\documentclass[aps,prl,reprint,groupedaddress]{revtex4}
\usepackage{mathrsfs}
\usepackage{amssymb}
\def\a{\alpha}
\def\b{\beta}
\def\t{\widetilde}

\def\p{\partial}  
  
\def\o{\over}
\def\p{\partial}
\def\l{\left}
\def\r{\right}

\def\U5{\tilde U_5}
\def\be{\begin{equation}}
\def\ee{\end{equation}}
\def\bea{\begin{eqnarray}}
\def\eea{\end{eqnarray}}
\def\f2{F^{\a\b}F_{\a\b}}
\def\16gc2{{16\pi G\over 3c^2}}
\def\trho{{\t\rho}}

\begin{document}

%\preprint{}

\title{Long-range scalar forces in five-dimensional general relativity}

% Paper written over the period May through June 2020, completed 20 June 2020
%
% based on research accomplished Dec 2019 to April 2020


\author{L.L.~Williams}
\email{willi@konfluence.org}
%\homepage[www.konfluence.org]{}
%\thanks{}
\affiliation{Konfluence Research Institute,\\ Manitou Springs, Colorado}


%\date{\today}
\date{20 June 2020}
\begin{abstract}
Kaluza first observed that the vacuum Einstein equations written in 5 dimensions (5D) reproduce exactly 4D general relativity and classical electrodynamics, when derivatives of the 5D metric with respect to the 5th coordinate are set to zero. The 15th component of the 5D metric is a 4D long-range scalar field, and the 4D limit emerges as the scalar field goes to one. Because the exact 4D Einstein and Maxwell equations are reproduced, a classical experimental test of the 5D framework has been elusive. Here we report new analysis, results, and force effects from the long-range Kaluza scalar field. The gravitational mass of a planet is a bare mass that is clothed with the mass-energy of the scalar field. We provide an identification and discussion of the scalar charge that couples to the scalar field. A new, third characteristic lengthscale for the electro-gravitic fields of a body of mass $M$ and charge $Q$ is identified, $\mu_0 Q^2/M$, to go along with the Reissner-Nordstrom lengthscales, $GM/c^2$ and $Q\sqrt{G\mu_0 /c^2}$.  These considerations reveal a strong electro-gravitic buoyancy force arising from a gravitational coupling between the electric charge of a body and the scalar field around a planet. At atomic scales, 5D covariance of the source terms requires that the electric, scalar, and gravitational forces all become proportional to electric charge. We discuss these results relative to foundational work by Dicke on long range scalar interactions and the Brans-Dicke scalar-tensor theory, including how interaction with the scalar implies a variable rest mass, and the energetics of the joint scalar-gravitational interaction. We discuss the problems of tuning the only available free parameter, the invariant 5D length element of the sources, to neutralize the scalar force and avoid the otherwise-large scalar force predictions. We conclude that large classical forces are a testable prediction of 5D general relativity sufficient to falsify the 5D hypothesis. Yet the emergence of a new physical interaction lengthscale is suggestive of further unexplored physics of the long range scalar field.

\end{abstract}

%\keywords{}

\maketitle

\section{I. INTRODUCTION}

Four years after his discovery of the field equations of general relativity, Einstein received a paper from Kaluza \cite{kal} showing that the field equations of general relativity, and the field equations of electromagnetism, behave as if the gravitational tensor field and the electromagnetic vector field are components of a 5-dimensional (5D) gravitational tensor field. Since a 5D tensor has 15 components, an additional long-range scalar field potential is implied. Standard 4D physics is recovered when this scalar potential goes to one, and that was Kaluza's original assumption. 

The 5D picture holds not only for the field equations, but for the equations of motion too, as Kaluza originally showed. When the geodesic equation is written in 5 dimensions, it is found to contain the standard 4D geodesic equation, along with the Lorentz force law of electromagnetism. As in the field equations, the 4D limit of the equations of motion is obtained when the scalar field goes to one, but there are otherwise additional long-range scalar force effects not in standard physics. 

We are careful not to confuse the ``Kaluza-Klein'' theory of compact dimensions with the strictly classical Kaluza theory of 5D general relativity and its long-range scalar field. After Kaluza's paper, field equations that properly incorporated the long-range scalar field were developed to various degrees by independent research groups: led by Einstein at Princeton (1930s-1940s), by Lichnerowicz in France (1940s), Scherrer in Switzerland (1940s), by Jordan in Germany (1950s), and by Dicke at Princeton (1960s).\cite{gon}, \cite{berg}

Progress on analysis of the field equations accelerated in the 1980s with the availability of English translations \cite{acf} of work by Thiry \cite{thry} under Lichnerowicz. During the 1980s and 1990s, significant work was done on the Kaluza theory, including a general solution for the 5D metric of a charged object \cite{cd}, analogous to the Schwarzschild and Reissner-Nordstrom solutions. 

The review by Gonner \cite{gon} shows that the equations of the Kaluza scalar field were developed by multiple European research groups in the mid-20th century, but none of those results were published in English language journals. The war caused further disruptions in the dissemination of results obtained by the European groups. The English-language Kaluza literature of the late 20th century (see Ref.~\cite{ow} for a review) contains variations in the field equations, and various authors make differing assumptions regarding the form of the Kaluza Lagrangian. The correct form of the Kaluza Lagrangian was obtained by Refs.~\cite{coq} and \cite{fri}, and their result was verified using tensor algebra software \cite{wil}. Ref.~\cite{fri} also obtains the correct curvature tensors, while Ref.~\cite{coq} has some minor errors in the 5D Ricci tensors.\cite{wil}

Another notable feature of the 5D ansatz introduced by Kaluza was the cylinder condition, that no field component depends on the fifth coordinate, and so its derivatives vanish. This was seen by Kaluza as the mathematical expression of the absence of a detectable fifth coordinate. 

In 1926, Klein \cite{kln} adapted the Kaluza 5D ansatz to quantum considerations, and hypothesized a compact, microscopic fifth dimension, thereby bringing Planck's constant and quantum considerations into a framework that is typically known as ``Kaluza-Klein". Klein's suggestion turned out to be a forced marriage of classical and quantum theory that did not satisfactorily describe 4D physics, but it profoundly influenced the direction of quantum field theory, making compact small dimensions an accepted part of physics, in the Kaluza theory \cite{eb}, and in higher-dimensional theories \cite{acf}. Today, many researchers reflexively think of the Kaluza fifth dimension as compact and microscopic.

It is important to bear in mind that this work is entirely classical, a treatment of general relativity in 5 dimensions. The entirety of the mathematics is simply to write general relativity in 5 dimensions instead of 4, and to set derivatives with respect to the 5th dimension constant. When this is done, 4D general relativity and classical electrodynamics are reproduced perfectly. The debate ensues about what this mathematics ``means". Does it mean the 5th dimension is ``real"? Is it compact? Is it microscopic? Fortunately, since we have force equations in this theory, the answer to these philosophical questions are irrelevant to testable predictions.

The results here cannot be constrained by particle-accelerator measurements of sub-atomic structure, just as they cannot constrain general relativity or the Maxwell equations. Indeed, the validity of a classical theory of charged particles is confined {\it only} to results independent of particle structure.\cite{ror} Certainly the Maxwell equations and the Einstein equations describe macroscopic fields whose underlying reality is quantum, yet their classical averages still yield testable predictions. So, too, do we apply the classical 5D theory, in the hopes that it can still yield testable predictions independent of atomic structure. Rohrlich \cite{ror} assures us that classical theories can yield valid descriptions of atomic systems in those cases where the result does not depend on assumptions of atomic structure, and where the results are convergent.

Even so, most of the references in the Kaluza literature adopt the traditional classical field equations and the cylinder condition, while opining in the text of a microscopic, compact fifth dimension. The math is still classical, but classical thinking is abandoned. What are naturally interpreted as classical long-range gravitational, electromagnetic, and scalar fields, are reinterpreted to be the ``$n=0$" modes of a fourier expansion of fields, on a manifold with a compact 5th dimension \cite{acf}. Ad hoc quantum considerations are laid on the classical theory, and Planck's constant emerges amid new free parameters. 

This seems to violate the spirit of a consistent classical theory of charged particles, as enunciated by Rohrlich. Correspondingly, we make no assumption that the fifth dimension is compact, since that is not necessitated by the Kaluza field equations or by the cylinder condition \cite{ow}. This work is not ``Kaluza-Klein'' theory, but the strictly classical Kaluza theory of 5D general relativity. The conceptual framework we adopt for the fixed mathematics is that the fifth dimension is open and macroscopic, like the other four of spacetime. No macroscopic classical experiment contradicts that assumption. Classical theories can only be tested in classical experiments. As we will see, the cylinder condition produces a non-trivial constant of the motion irrelevant to compact dimensions.

The unique nature of the fields in 5D general relativity and their mutual coupling are revealed in the Kaluza Lagrangian:
\begin{equation}
\label{Lag}
\mathscr{L} = g^{1/2}\left[{c^4\phi\o 16\pi G}\ g^{\a\b} R_{\a\b} -{\phi^3\o 4\mu_0}\ g^{\a\mu} g^{\b\nu} F_{\a\b} F_{\mu\nu} \right]
\end{equation}
where $G$ is the gravitational constant, $\mu_0$ is the permeability of free space, $R_{\mu\nu}$ is the Ricci tensor, $F_{\mu\nu}\equiv \p_\mu A_\nu - \p_\nu A_\mu$ is the electromagnetic field strength tensor, $g$ is the determinant of $g_{\mu\nu}$, and the gravitational field equations are obtained from variation with respect to $g^{\mu\nu}$. $\phi$ is a new scalar field necessitated by the 5D metric hypothesis. We can call this the long-range Kaluza scalar field. This form is obtained by Refs.~\cite{coq} and \cite{fri}, and verified by Ref.~\cite{wil} using tensor algebra software. It is clear that the 4D limit of the theory occurs when $\phi\rightarrow 1$.%, consistent with (\ref{ksi}).

The scalar-electromagnetic couplings make the Kaluza theory unique among scalar-tensor theories, and markedly different from Brans-Dicke theory. The Kaluza Lagrangian (\ref{Lag}) contains aspects of the Brans-Dicke scalar-tensor Lagrangian \cite{bd}, \cite{wbg}. In each case, the scalar field can be viewed as a variable gravitational constant. The Kaluza Lagrangian also contains aspects of the Bekenstein scalar-electromagnetic theory \cite{bk}, with a variable vacuum permittivity, although Bekenstein did not start from a Lagrangian, and spoke in terms of a variable fine structure constant. 

Treatment of the 5D source term varies in the literature. Kaluza and subsequent authors typically assume weak specific charges in the source terms, to avoid conceptual issues or deviations from standard physics at high specific charges. By specific charge, we mean the ratio of electric charge to mass of a body. In this classical theory, charge is a specific quantity that can go smoothly to zero, and the charge carriers are not quantized, consistent with the demand that the classical predictions be independent of particle structure.

Although the 5D geodesic equation has been long known and well-studied \cite{gk}, \cite{fri}, \cite{wpl}, \cite{cd}, the corresponding energy-momentum tensor was obtained only recently \cite{wil2}. General covariance, and the requirement that the source term be a 5D tensor to match the 5D Einstein tensor, is essential to establishing the correct 5D energy-momentum tensor corresponding to the 5D geodesic equation. When 5D covariance of the source terms is established, unique saturation effects emerge in the source terms. The saturation effects are such that the coupling of a test body to either the gravitational, electric, or scalar forces can vary with its specific charge.

In this work, we examine the charge-specific effects in the source terms and their implications for the coupled equations of gravity, the electric force, and the scalar force. We recover the intriguing result, originally described by Dicke \cite{dke}, \cite{dke2}, that the scalar field coexists with the Newtonian gravitational field. Furthermore, the mass-energy of the Kaluza scalar field contributes to the total spacetime curvature we measure through Kepler's laws. 

We also find that the Kaluza theory implies an electro-gravitic buoyancy effect around planet-sized masses. Unlike Brans-Dicke scalar-tensor theory, the coupling to the Kaluza scalar field is not through mass-energy, but through the Kaluza scalar charge, which is proportional to electric charge. The sign of the scalar field is opposite to the Newtonian gravitational potential arising from the perturbation of the time-time component of the metric. Because the scalar charge is independent of the sign of charge, any electrically charged object should feel an upward, counter-gravitational force. The predicted magnitude for the effect is large, perhaps too large to be true. Yet if so, it would provide the first experimental falsification of the Kaluza hypothesis. Until now, the theory has always reproduced known classical physics in the limit of a constant scalar field, and the Kaluza theory has limited freedom of parameters.

We find that the saturation effects in the source terms act to alter the nature of the scalar coupling in high specific charge environments, so that the Kaluza scalar field may masquerade as the electric force in the parameter regimes of atomic systems. Furthermore, it appears that the scalar potential goes to zero for point particles, suppressing the scalar force altogether for atomic systems.

In the following development, we consider the covariant field equations with sources, and identify gravitational, electric, and scalar charges. From the general field equations in the presence of sources, we provide static, spherically-symmetric solutions for the scalar, electric, and gravitational potentials around charged, massive bodies. Three limits in the solutions are encountered for different values of specific charge: neutral, weak charge states, and strong charge states. From the potentials and the charges, the scalar, electric, and gravitational forces between charged bodies are established. 

A new electro-gravitic lengthscale is discovered, which complements the known electric and gravitational lengthscale in the Reissner-Nordstrom metric. Gravitating mass is found to be clothed by a scalar field bound to the bare mass. A unique electro-gravitic buoyancy effect is discovered in the bound scalar field. In light of large, undiscovered forces implied by the mathematics, we discuss how the scalar effects might be tuned to zero. The hierarchy of lengthscales, from terrestrial to cosmological, is discussed. Saturation effects for ultra-high specific charge states characteristic of atomic system appear to allow the scalar force and gravitational force to masquerade as the electric force, but both go to zero for point particles. All results are compared to work by Dicke on long range scalar fields, including energy considerations and the effect of long-range scalar fields on mass.

%%%%%%%%%%%%%%%%%%%%%%%%%%%%%%%%%
%%%%%%%%%%%%%%%%%%%%%%%%%%%%%%
\section{II. SOURCES AND CHARGES}

In this section we provide an assessment of source terms in the Kaluza field equations. We relate the long-studied 5D geodesic equation to new considerations regarding the 5D covariance of source terms in the field equations. The gravitational, electromagnetic, and scalar charges are identified, along with their dependence on the long-range Kaluza scalar field.

\subsection{A. Covariant field sources}
%%%%%%%%%%%%%%%%%%%%%%%%%%%%%%%%%%
%%%%%%%%%%%%%%%%%%%%%%%%%%%%%%%%%%

The covariant 5D Einstein equations, for a 5D energy-momentum tensor of a cold fluid, corresponding to the usual 5D geodesic equation, are given by \cite{wil2}:
\begin{equation}
{\t G}_{ab} = {8\pi \t G\over c^3}
{\t\rho\over {\t g}^{1/2}}{{\t U}_a {\t U}_b\over (dt/ds)}
\label{5dee}
\end{equation}
where $\t\rho$ is a rest mass density per volume per fifth coordinate, ${\t G}_{ab}$ is a 5D Einstein tensor for a 5D metric ${\t g}_{ab}$, $\t G$ is a 5D gravitational constant, and $c$ is the speed of light. Small roman indices range over the 5 dimensions. The form of (\ref{5dee}) assigns the time coordinate $t$ as the independent variable.

It is standard to assign the components of the 5D metric to the 4D metric $g_{\mu\nu}$, the electromagnetic potential 4-vector $A^\mu$, and a scalar potential $\phi$, as follows:
\begin{equation}
\label{metric}
\t g_{\mu\nu} = g_{\mu\nu} + \phi^2 k^2 A_\mu A_\nu \ ,  \quad
\t g_{5\nu} = \phi^2 k A_\nu \ , \quad
\t g_{55} = \phi^2
\end{equation}
Here, greek indices range over the 4 dimensions of spacetime. The index $5$ denotes the fifth dimension. Since ${\t g}_{ab}{\t g}^{bc} = \delta^c_a$, the inverse metric is given by:
\begin{equation}
\label{invmet}
\t g^{\mu\nu} = g^{\mu\nu}  \ ,  \quad
\t g^{5\nu} = - kA^\nu \ , \quad
\t g^{55} = k^2 A^2 + {1/\phi^2}
\end{equation}

The constant $k$ is the characteristic electro-gravitic scale parameter of the Kaluza theory, given in MKS units as \cite{wil2}:
\begin{equation}kc \equiv \sqrt{16\pi G \epsilon_0} = \sqrt{16\pi G/\mu_0 c^2} \simeq 1.7\times 10^{-10}\ {\rm C/kg}
\label{k}
\end{equation} 
It is closely related to the ADM mass \cite{adm}.

%%%%%%%%%%%%%%%%%%%%%%%%%%%%%%%%%%
%%%%%%%%%%%%%%%%%%%%%%%%%%%%%%
\subsection{B. Invariant constraint on sources}
%%%%%%%%%%%%%%%%%%%%%%%%%%%%%%
%%%%%%%%%%%%%%%%%%%%%%%%%%%%%%

The 5D proper velocity is defined by 
\begin{equation}
{\t U}^a \equiv {dx^a\over ds}
\end{equation}
where the 5D length element is given by
\begin{eqnarray}
\label{5dle}
\varepsilon_a{\t a}^2 ds^2 &=& {\t g}_{ab}dx^a dx^b\cr &=& g_{\mu\nu}dx^\mu dx^\nu + \varepsilon_\phi\phi^2 (dx^5 + kA_\nu dx^\nu)^2
\end{eqnarray}
In the 5D length element (\ref{5dle}), the parameters $\varepsilon_a,\varepsilon_\phi = \pm 1$. They account for the fact that the 5D hypothesis does not fix the sign, timelike or spacelike, of the 5D length element, $\varepsilon_a$, or of the fifth metric component, $\varepsilon_\phi$. 

Variation of the $\varepsilon_a$ and $\varepsilon_\phi$ can lead to differing saturation effects compared to what we will report. We find that a non-imaginary mass requires that $\varepsilon_a = +1$ and $\varepsilon_\phi = +1$, so that both the 5D length element and the signature of the fifth dimension in the metric are timelike. As mentioned above, we avoid imaginary-mass solutions because of their absence from planetary and laboratory physics.

Results in the Kaluza literature often indicate the signature of the fifth dimension is spacelike. However, the form of the Kaluza Lagrangian (\ref{Lag}) seems to suggest that the signature of the fifth coordinate is timelike \cite{wil}. An experimental investigation \cite{tw} also lends support to that conclusion, by testing for the existence of a rest-frame for motion along the fifth coordinate. We therefore adopt $\varepsilon_a=\varepsilon_\phi = +1$ in the following, but the alternatives are easily inferred from results reported here.

The 5D length element (\ref{5dle}) implies an invariant length $\t a$ of the five-vector ${\t U}^a$:
\begin{equation}
\label{a2}
{\t a}^2 = {\t g}_{ab}{\t U}^a {\t U}^b = \left( cd\tau\over d s \right)^2 + \phi^2 \left({\t U}^5 + kA_\nu {\t U}^\mu\right)^2
\end{equation}
where we used the definition of 4D proper time, $c^2 d\tau^2 \equiv g_{\mu\nu}dx^\mu dx^\nu$.
$\t a$ is a free parameter in the 5D length element that must be fixed by correspondence to known physics. In fact, this is the only free numerical parameter in 5D general relativity. As with the $\varepsilon_a$ and $\varepsilon_\phi$, differing choices of $\t a$ can lead to different saturation effects in the couplings than reported here.

Applying a standard result of general relativity, we see that the absence of a dependence of the metric on the fifth coordinate, $\p_5 {\t g}_{ab}=0$, implies that that the fifth covariant component of $\t U^a$ is a 5D constant of the motion:
\begin{equation}
\label{U5}
{\t U}_5 = {\t g}_{5b} {\t U}^b = 
\phi^2 ({\t U}^5 + kA_\nu {\t U}^\nu ) = {\rm constant}
\end{equation}
The cylinder condition implies a non-trivial constant of the motion, and this reinforces our choice to adopt a classical perspective and treat the cylinder condition as an imposed boundary condition, somewhat akin to a time-independent boundary condition that would imply a conserved energy. 

The constant (\ref{U5}) allows us to simplify (\ref{a2}):
\begin{equation}
\label{a2u}
{\t a}^2 = \left( cd\tau\over d s \right)^2 + {{\t U}^2_5 \over \phi^2}
\end{equation}

Except for (\ref{5dee}) and (\ref{k}), the previous equations are standard in the classical Kaluza literature. At this point, we turn toward new results by noting first from (\ref{Lag}) that $\phi\rightarrow 1$ in the 4D limit. Therefore we contemplate a Newtonian-style perturbation expansion of $\phi$ such that
\begin{equation}
\label{ksi}
\phi \simeq 1 + \xi + O(\xi^2)\ ,\ \xi\ll 1
\end{equation}
We will verify at the end that $\xi\ll 1$. 

Let us now fix the constant $\t a$ from (\ref{a2u}). In the limit that ${\t U}_5 \rightarrow 0$, then $cd\tau /ds \rightarrow 1$. Similarly, we know that asymptotically from (\ref{ksi}), $\phi\rightarrow 1$. Therefore
\begin{equation}
\label{a2c}
{\t a}^2 \equiv 1 + {\t U}^2_5
\end{equation}
We see that $(1+{\t U}^2_5)ds^2$ plays the same role in 5D as $c^2d\tau$ does in 4D, an invariant length element, with a characteristic velocity.

Now let us use (\ref{ksi}) and (\ref{a2c}) to rewrite (\ref{a2u}):
\begin{eqnarray}
\label{emass}
\left( cd\tau\over d s \right)^2 &=& 1 + {\t U}_5^2 (1-\phi^{-2}) = 1 + 2 \xi {\t U}_5^2  + O(\xi^2) \cr
&\simeq& 1 + 2 \xi {\t U}_5^2
\end{eqnarray}
This is a new expression in Kaluza theory. It acts as a coupling coefficient, and manifests the aforementioned saturation effects.

%%%%%%%%%%%%%%%%%%%%%%%%%%%%%%
%%%%%%%%%%%%%%%%%%%%%%%%%%%%%%
\subsection{C. Identification of charges and fields}
%%%%%%%%%%%%%%%%%%%%%%%%%%%%%
%%%%%%%%%%%%%%%%%%%%%%%%%%%%%%

\begin{widetext}
Although the 5D geodesic equation is standard in the Kaluza literature, let us revisit it from the 5D Bianchi identities applied to (\ref{5dee}):
\begin{equation}
\label{5db}
{\t\nabla}_a \left ( {\t\rho\over {\t g}^{1/2}} {{\t U}^a {\t U}^b\over (dt/ds)} \right ) = 
{\t U}^b{\t\nabla}_a \left( {\t\rho\over {\t g}^{1/2}} {\t U}^a \right)
+ {\t\rho\over {\t g}^{1/2}} {\t U}^a {\t\nabla}_a {\t U}^b = 0
\end{equation}
When particles are conserved in 5D, the first term on the RHS is zero, and we obtain the usual 5D geodesic equation, ${\t U}^a {\t\nabla}_a {\t U}^b = 0$. The geodesic equation is an expression of 5D conservation of energy-momentum independent of $\trho$.

The 5D geodesic equation can be cast in terms of 4D quantities \cite{wil2}
\begin{equation}
\label{4Deom}
{d\tau\over ds}\left({dU^\nu\o d\tau} + {\Gamma}^\nu_{\a\b} {U}^\a U^\b \right) = 
k {\t U}_5 g^{\nu\mu}F_{\mu\a}U^\a 
+ {\t U}_5^2 \left({ds\over d\tau}\right){(\p_\a \phi)\over\phi^3} \l [ g^{\nu\a} - {U^\nu U^\a \over c^2 } \r ]
\end{equation}
\end{widetext}
where the 4D proper velocity of a particle is
\be
U^\mu \equiv {dx^\mu\o d\tau}\ .
\ee

Equation (\ref{4Deom}) has been studied by various researchers, including \cite{fri}, \cite{gk}, \cite{wpl}. The term in brackets on the RHS of (\ref{4Deom}) that is quadratic in $U^\a$ arises from the transformation of derivatives with respect to $s$, to derivatives with respect to $\tau$, and using (\ref{U5}). This is indeed the form expected for a scalar field force.\cite{jksn} 

The term linear in ${\t U}_5$ in (\ref{4Deom}) must be identified with electric charge, $Q$, of a body of rest mass $M$, to correspond with the Lorentz force law. The coefficient of the gravitational terms in (\ref{4Deom}) is identified with mass, and the coefficient of the scalar field term is identified as the Kaluza scalar charge. 

We therefore identify the three charges associated with the three forces: mass for gravity, electric charge for electromagnetism, and a new scalar charge for the scalar force. Multiplying (\ref{4Deom}) through by $Mc$ and using (\ref{emass}):
\begin{equation}
\label{mass}
M{cd\tau\over ds} \simeq M \sqrt{1 + 2 \xi {\t U_5}^2} \equiv {\t M} \quad\longrightarrow \quad{\rm mass}
\end{equation}

\begin{equation}
\label{echg}
Mck {\t U}_5 \equiv Q \quad\longrightarrow \quad{\rm electric\ charge}
\end{equation}

\bea
\label{schg}
 M c{{\t U}_5^2} {ds\over d\tau} &\simeq&
{Mc^2 {\t U}^2_5 \over \sqrt{1 + 2 \xi {\t U}_5^2}}
 \equiv {\t S} \longrightarrow {\rm scalar\ charge}\cr
&=& {Q^2\o {\t M}k^2}
\eea
The assignments of mass (\ref{mass}) and charge (\ref{echg}) are common in the Kaluza literature, e.g., \cite{fri}, \cite{gk}. The expression for scalar charge is more variable in the Kaluza literature, and (\ref{schg}) is a new result, as is the $\xi$ dependence of (\ref{mass}). 

The expression for mass (\ref{mass}) implied by (\ref{emass}) and (\ref{echg}) is seen to have a peculiar dependence on electric charge and the long-range scalar field $\phi$. This is to be expected. It is characteristic of long range scalar fields that, if they interact with a particle, the particle mass must be a function of the scalar field.\cite{dke},\cite{dke2} In this theory, the scalar field brings to life a mass variation for charged bodies. However, since $\xi\ll 1$, the variation is small for all charged systems. No macroscopic experiment should show this effect.

An effective mass of the form (\ref{mass}) is common in the Kaluza literature, but with variation. The form used in (\ref{mass}) matches closely Ref.~\cite{gk}, and is similar in nature to that in Ref.~\cite{fri}. Ref.~\cite{coq} uses an entirely different form. The work presented here is unique in its parameterization of the Kaluza scalar field in terms of $\xi$, via (\ref{ksi}), and will lead to unique results. Previous research on the Kaluza scalar field has not considered the implications of a Newtonian-like perturbation.

The form of (\ref{mass}) is positive-definite, due to our choice of $\epsilon_a$ in (\ref{5dle}). It ascribes an increase in mass to interaction with the scalar field. Choosing the opposite sign of $\epsilon_a$ would lead to a decrease in mass, but also potentially to imaginary mass. The concept of imaginary mass has a place in physis, but we wish to avoid it here on the grounds that planetary masses are understood to be real numbers and there is no accepted interpretation of the gravitational field of an imaginary mass.

Electric charge is identified in (\ref{echg}) with a 5D constant of the motion, and is functionally invariant for all charge states. The constant $ck$ (\ref{k}) forms a characteristic charge-to-mass ratio. It is better known as the ADM mass \cite{adm} when combined with the quantum of electric charge. The preceding analysis may indicate why no particle with the ADM mass exists. It is not a breakdown of the classical theory, but is a misappropriation of an intrinsic quantity, a universal charge-to-mass ratio.

Note from (\ref{U5}) that there is a term of $\t U_5$ that depends on particle 4-velocity. Therefore, the 5D invariant electric charge $Q\propto \t U_5$ can be understood as a sort of canonical electric charge, analogous to the canonical momentum of a particle in an electromagnetic field. However, the motional charge is proportional to $kA_\mu$, which is typically very small, perhaps too small to measure.

The scalar charge expression (\ref{schg}) takes more variable forms in the literature, because there is no mapping to known physics of the Kaluza scalar force. Its form was very much an open question in the monopole solution by Ref.~\cite{cd}. The Kaluza scalar field perturbation $\xi\ll 1$, and so for many electrically-charged objects, the scalar charge $\propto Q^2/M$. For uncharged objects, both electric charge and scalar charge need not vanish, due to the ``canonical" or induced charges in (\ref{U5}) $\propto kA_\mu U^\mu$. 

For highly charged objects, when ${\t U_5} \gg \xi^{-1}$, the scalar charge saturates to a value linear in Q:
\begin{equation}
\label{schgx}
{\rm saturated\ scalar\ charge}\quad\longrightarrow\quad {Mc^2 {\t U_5}\over \sqrt{2\xi}}, \quad 
{\t U_5}\gg \xi^{-1}
\end{equation} 
That is, the scalar coupling becomes proportional to electric charge for highly-specific-charge objects such as elementary particles. This means the scalar force merges with the electric force, and the scalar force masquerading Dicke described for gravitational fields is seen here for electric fields as well. A similar effect is also seen for the saturated mass. 
\begin{equation}
\label{massx}
{\rm saturated\ mass}\quad\longrightarrow\quad {M {\t U_5} \sqrt{2\xi}}, \quad 
{\t U_5}\gg \xi^{-1}
\end{equation} 

The implication of these results will be developed in greater detail in the following.

%%%%%%%%%%%%%%%%%%%%%%%%%%%%%%%%%%%%%%%%%%%%%
%%%%%%%%%%%%%%%%%%%%%%%%%%%%%%%%%%%%%%%%%%
\subsection{D. Newton's Third Law}
%%%%%%%%%%%%%%%%%%%%%%%%%%%%%%%%%%%%%%%%%%%%%
%%%%%%%%%%%%%%%%%%%%%%%%%%%%%%%%%%%%%%%%%%
According to the equation of motion (\ref{4Deom}), the force on a body is the product of the gradient of the potential and its corresponding charge. We expect that Newton's Third Law is obeyed, so that the product of charge 1 in the field of charge 2 is the same as the product of charge 2 in the field of charge 1. 

The identity is enforced by the form of (\ref{5dee}). It is seen that the sources can be understood as the flux of an energy-momentum-charge 5-vector, just as the energy-momentum tensor for 4D dust can be understood as the flux of an energy-momentum 4-vector. In the latter case, the energy corresponds to a momentum in time, in that it arises from a change in the time coordinate, just as momentum arises from a change in the space coordinate. 

The spacetime flux of the 5th component of the 5-velocity corresponds to the electric current 4-vector. The change along the 5th coordinate of the 5th component of the 5-velocity corresponds to scalar charge. The 5th component of the 5-velocity corresponds to the energy bound into charge, just as the time component of the 5-velocity describes the energy bound into rest mass. 

So we can rephrase the identifications (\ref{mass}), (\ref{echg}), and (\ref{schg}) by referring directly to (\ref{5dee}) and (\ref{4Deom}):
\be
\label{3mass}
{\rm gravitational\ source\ current}\quad\longrightarrow\quad {\t\rho}\left({cd\tau\o ds}\right) U_\mu U_\nu
\ee
\be
\label{3charge}
{\rm electromagnetic\ source\ current}\quad\longrightarrow\quad {\t\rho} {\t U}_5 U_\mu
\ee
\be
\label{3scalar}
{\rm scalar\ source\ current}\quad\longrightarrow\quad {\t\rho} {\t U}_5^2
\left({ds\o cd\tau}\right)
\ee

These sources appear both in the 5D Einstein equations and the equations of motion. Therefore, the gravitational, electromagnetic, and scalar forces between particles and the fields they source act symmetrically, so that the force exerted on another particle through its field is equal to the force that particle experiences from the field of the other particle. This satisfies Newton's Third Law individually for each of the 3 forces.

%%%%%%%%%%%%%%%%%%%%%%%%%%%%%%
%%%%%%%%%%%%%%%%%%%%%%%%%%%%%%
\section{III. FIELDS OF MATTER SOURCES}

In this section we solve the general field equations for time-independent, spherically-symmetric solutions. The Kaluza scalar field is treated as a small perturbation, like the gravitational field. The gravitational, electric, and scalar fields are obtained for non-relativistic sources. Due to the dependence of scalar charge and mass on electric charge, 3 limits of specific charge are investigated: neutral matter, achievable laboratory specific charge, and specific charge characteristic of atomic systems.

\subsection{A. General field equations}
%%%%%%%%%%%%%%%%%%%%%%%%%%%%%%
%%%%%%%%%%%%%%%%%%%%%%%%%%%%%%

The general field equations for the Lagrangian (\ref{Lag}), for the covariant 5D source term as in (\ref{5dee}), are provided by Ref.~\cite{wil2}. We recapitulate those equations here without derivation.

The gravitational field equations for $g_{\mu\nu}$ are obtained from the components ${\t G}_{\mu\nu}$ of (\ref{5dee}):
\begin{eqnarray}
\label{kee}
G_{\mu\nu} =  \phi^{-1} T^\phi_{\mu\nu}
&+& {8\pi G\o\mu_0 c^4}\phi^2 T_{\mu\nu}^{EM} \\ &+& 
{8\pi G\o c^3\phi}{d\tau\o ds}{\t\rho\o g^{1/2}} g_{\mu\a} {dx^\a\o dt} U_\nu \nonumber
\end{eqnarray}
where
\be
\label{tphi}
T^\phi_{\mu\nu} \equiv \nabla_\mu \nabla_\nu \phi - g_{\mu\nu} \nabla_\a \nabla^\a \phi
\ee
is the Kaluza scalar field energy-momentum tensor, and where
\be
T_{\mu\nu}^{EM} \equiv g^{\a\b} F_{\mu\a} F_{\nu\b} - {1\o 4} g_{\mu\nu} F_{\a\b} F^{\a\b}
\ee
is the electromagnetic energy-momentum tensor. The Kaluza modification to the Einstein equations comes in the scalar field energy-momentum, in the electromagnetic energy-momentum, and in the term in $d\tau /ds$ in the material energy-momentum. The Kaluza scalar field behaves like a variable gravitational constant, as in the Brans-Dicke scalar-tensor theory, {\it except} with respect to its coupling to electromagnetic energy-momentum.

The electromagnetic field equations for $A^\mu$ are obtained from the components ${\t G}_{5\nu}$ of (\ref{5dee}):
\begin{equation}
\label{max}
\nabla_\nu (\phi^3 F^{\nu\mu}) = \mu_0{\t\rho\o g^{1/2}}  kc {\t U}_5{dx^\mu \o dt} 
\end{equation}
%2May20 verified phi^3 on LHS includes phi from source term
The Kaluza modification to the Maxwell equations comes in the scalar field, which acts as a variable dielectric constant, similar to the Bekenstein theory \cite{bk}. The quantity $\t\rho kc\t U_5$ is immediately identified with the electric charge density, reproducing the expected form of the Maxwell equations. The quantity $\phi^3 F_{\mu\nu}$ emerges as an invariant under conformal transformations.\cite{coq}

The scalar field equation for $\phi$ is obtained from the component ${\t G}_{55}$ of (\ref{5dee}):
\begin{equation}
\label{ksf}
\phi^2 \left ( {3\o 4} \phi^2 k^2 F_{\a\b}F^{\a\b} - R \right ) =
{16\pi G\o c^3\phi}{ds\o dt}{\t\rho\o g^{1/2}}{\t U}_5^2
\end{equation}
This equation is new to physics. Therefore our inferences about it rely on its mathematical emergence alongside the electromagnetic and gravitational fields, which we can constrain with known physics. 

The scalar curvature enters (\ref{ksf}), and is given by the trace of (\ref{kee}), as usual:
\begin{equation}
\label{trace}
R = {3\o \phi} {\nabla_\a \nabla^\a \phi} - {8\pi G\o c\phi}{d\tau\o ds}{\t\rho \o g^{1/2}} {d\tau\o dt}
\end{equation}
where the sign of the matter term depends on the 4D metric signature. In this work, we adopt $(+,-,-,-)$, so that the trace of the matter term is positive. 

Because the scalar curvature involves second order derivatives of $\phi$, the equation (\ref{ksf}) for $\phi$ becomes dynamical, even though $\phi$ enters only algebraically in the Lagrangian (\ref{Lag}). Therefore we can rewrite (\ref{ksf}) using (\ref{trace}) to explicitly show the scalar field derivatives:
\bea
\label{ksffull}
-3{\nabla_\a \nabla^\a \phi} &=&  \mu_0 k^2 c {\trho \o g^{1/2}} {{\t U}_5^2\o\phi^2}{ds\o dt}
- {8\pi G\o c}{d\tau\o ds}{\t\rho \o g^{1/2}} {d\tau\o dt} \cr
&-&  {3\o 4} \phi^3 k^2 F_{\a\b}F^{\a\b}
\eea

From (\ref{trace}) and (\ref{ksf}), we see that in the absence of electric charge and electromagnetic fields and sources, the Kaluza scalar field will act to neutralize the scalar curvature by enforcing $R=0$ against whatever sources of matter exist in spacetime. The scalar field will play a role in the total energy budget of spacetime, and in this way masquerade as gravity \cite{bd}, as we will see shortly.

Also note that charged matter and neutral matter enter as sources for $\phi$ with opposite sign.

The field equations and the sources obey a constraint from the 5D trace of the 5D field equations, (\ref{5dee}), using (\ref{a2}) and (\ref{a2c}):
\be
\label{5traceEMT}
{\t g}_{ab}{\t U}^a {\t U}^b = 1 + {\t U}^2_5
\ee

%\begin{comment}
An alternative equation for $\phi$ can be obtained that provides some insight by explicitly observing the constraint of the invariant length element (\ref{5dle}). Let us consider the 5D trace of (\ref{5dee}), using (\ref{5traceEMT}):
\begin{equation}
\label{5trace}
-{3\over 2} {\t R} = {8\pi G\over c^3} { {\t \rho} (1 + {\t U}^2_5) \over \phi g^{1/2}}{ds\over dt}
\end{equation}
The factor $3/2$ arises because $\t g_{ab} \t g^{ab} = 5$.

5D connections and curvature tensors for the 5D metric (\ref{metric}) are tabulated by Ref.~\cite{wil}. The 5D scalar curvature is given by:
\begin{equation}
\label{5scalar}
\t R = R - {1\over 4}k^2 \phi^2 F^{\a\b} F_{\a\b} - {3\o \phi} {\nabla_\a \nabla^\a \phi}
\end{equation}
Taken together, (\ref{5trace}) and (\ref{5scalar}) have the same information as (\ref{ksffull}). This demonstrates why the scalar charge can be related to the electric and gravitational charges.

%%%%%%%%%%%%%%%%%%%%%%%%%%%%%%
%%%%%%%%%%%%%%%%%%%%%%%%%%%%%%
\subsection{B. Constraints on solutions}
%%%%%%%%%%%%%%%%%%%%%%%%%%%%%%
%%%%%%%%%%%%%%%%%%%%%%%%%%%%%%

The general field equations given above for gravitational, electromagnetic, and scalar field are solved now under a series of typical constraints, which we enumerate here for convenience:

\begin{enumerate}
\item{spherical symmetry: fields depend spatially only on a radial coordinate, $r$}
\item{time independent: time derivatives vanish}
\item{magnetic fields vanish, so $F^{\a\b} F_{\a\b} = -2E^2/c^2$}
\item{sources are at rest}
\item{test particle speeds $v\ll c$}
\item{consider weak perturbations of the gravitational field, $g_{\mu\nu} \simeq \eta_{\mu\nu} + h_{\mu\nu},\ h_{\mu\nu}\ll 1$. With regard to gravity, this is a Newtonian limit}
\item{consider weak perturbations of the scalar field, such that $\phi \simeq 1 + \xi,\ \xi\ll 1$}
\item{the metric signature is $(+,-,-,-)$}
\end{enumerate}

Constraint no.5 implies $dt/d\tau =1$. It also implies that the gravitational force term in the geodesic equation (\ref{4Deom}) is dominated by $\Gamma^\nu_{tt}$, the time-time components of the connection. This is of course the component of relativistic gravity that accounts for the Newtonian limit.

Constraints nos. 2, 6, 8 imply $\nabla_\mu \nabla^\mu = -\nabla^2$, where $\nabla^2$ is the ordinary 3-space Laplacian operator.

We will apply the field equations at planetary and atomic scales. Our application to atomic scales will follow the prescription for predictions which do not rely on atomic structure, and which are well-behaved when an artifical size parameter goes to zero \cite{ror}.

Under the assumptions above, our task is to solve the coupled gravitational, electromagnetic, and scalar field equations reduces to solving coupled equations for 3 potentials:
\begin{itemize}
\item{the scalar perturbation $\psi$ of the time-time component of the metric, $g_{tt} \simeq 1 + \psi$}
\item{a radial electric field ${\bf E}(r)$, given by the radial gradient of the Coulomb potential}
\item{the perturbation $\xi$ of the Kaluza scalar field, $\phi \simeq 1 + \xi$}
\end{itemize}

We will conduct our analysis by investigating first-integrals of the field equations. This will allow us to investigate properties of the fields which are independent of particle structure. We examine volume integrals of the sources, and relate our findings back to Newton's law of gravity and Coulomb's law.

%%%%%%%%%%%%%%%%%%%%%%
%%%%%%%%%%%%%%%%%%%%%%%%%%%%%%%
\subsection{C. Scalar field of a planet}
%%%%%%%%%%%%%%%%%%%%%%%%%%
%%%%%%%%%%%%%%%%%%%%%%%%%

Now we consider the simplest case of an electrically-neutral body of mass $M$, anticipated to be of planetary magnitude. This corresponds to the case $\t U_5 =0$. Electric charge and electric fields vanish, so the problem reduces to two unknowns: the metric perturbation, $\psi$, and the scalar field perturbation, $\xi$. In this case, the mass (\ref{mass}) reduces to the rest mass $M$, and the electric and scalar charges (\ref{echg}) and (\ref{schg}) are zero.

We make a typical linearized expansion about the Minkowski metric, but in spherical coordinates. Recall that the diagonal components of the Minkowski metric $\eta_{\mu\nu}$ are $(c^2,-1,-r^2,-r^2\sin^2\theta)$, so that in this linearization, $\eta_{\mu\nu}$ is not constant.

As usual, the time-time component of the metric, $g_{tt}\simeq 1 + \psi$, where $\psi\ll 1$. The other components of the metric will also have perturbations, and we choose to work in standard form of the static, isotropic metric. In this form, there is no perturbation to the angular components of the Minkowski metric. The perturbations are only to the components $\eta_{tt}$ and $\eta_{rr}$.\cite{wbg2} Without solving the field equations for the $rr$ perturbation, we can safely assume it will be of the same order of magnitude as $\psi$. Therefore, we can write the perturbed determinant of the metric:
\be
g^{1/2} = r^2 \sin\theta (1 + \psi) +  \mathscr{O}(\psi^2)
\ee

Let us start with a general result which will be of use throughout these calculations. We write the Ricci tensor suggestively in terms of a divergence:
\bea
\label{eint}
-R_{\mu\nu} &\equiv& -R^\a_{\mu\a\nu} = \p_\nu \Gamma^\a_{\mu\a} - \p_\a \Gamma^\a_{\mu\nu} + 
\Gamma^\b_{\mu\a}\Gamma^\a_{\b\nu} - \Gamma^\b_{\mu\nu} \Gamma^\a_{\a\b} \cr
&=& -{\p_\a (g^{1/2} \Gamma^\a_{\mu\nu})\o g^{1/2} }+ \p_\mu \p_\nu \ln g^{1/2} + 
\Gamma^\b_{\mu\a}\Gamma^\a_{\b\nu}
\eea
where the second equality follows from a common identity for $\Gamma^\a_{\a\b}$ in terms of the determinant $g$ of the metric. The first term is then the covariant divergence of $\Gamma^\a_{\mu\nu}$, as if it were a set of ten 4-vectors, each of index $\mu\nu$. Then we note that only the components $\Gamma^\mu_{tt}$ enter Newtonian dynamics. Therefore, we need only evaluate $R_{tt}$ to get an equation for $\psi$.

For $R_{tt}$, the last two terms on the RHS of (\ref{eint}) vanish. The term in $g^{1/2}$ vanishes with the time derivatives. We assert, without showing here, that the term quadratic in the connections is of second order in the metric perturbations, $\Gamma^\b_{t\a}\Gamma^\a_{\b t} = 0 + \mathscr{O}(\psi^2)$. Therefore
\be
\label{r-tt}
R_{tt} =  g^{-1/2}{\p_i (g^{1/2} \Gamma^i_{tt}) } + \mathscr{O}(\psi^2)  \simeq {\bf\nabla}\cdot {\bf \Gamma}_{tt}
\ee
where we have illustrated our transformation with the notation of a 3-space divergence, as in Gauss's Law. It is clear, then, that a spherical volume integral of $R_{tt}$ will pick off only the radial component $\Gamma^r_{tt}$ of ${ \Gamma}^i_{tt}$. Therefore
\be
\label{rtt}
\int R_{tt} g^{1/2} d^3x \simeq
\int  \p_i (g^{1/2} \Gamma^i_{tt})d^3x = \oint \Gamma^r_{tt} r^2 d\Omega  
\ee

Let us then evaluate this component for the perturbed metric to find:
\be
\label{grtt}
\Gamma^r_{tt} = {1\o 2} \p_r \psi + \mathscr{O}(\psi^2)
\ee

Before proceeding to the source terms in the gravitational field equations, let us now turn to the scalar field equation. Under our constraints, the scalar field equation (\ref{ksf}) reduces to $R=0$. That is, the Kaluza scalar field has the peculiar quality of acting to neutralize the scalar curvature $R$ in spacetime arising from the presence of uncharged matter. 

Recall that we are linearizing such that $\phi \simeq 1 + \xi$. From (\ref{trace}), we obtain:
\be
\label{ksir}
\nabla^2 \xi = {1\o g^{1/2}}\p_i (g^{1/2} \p^i \xi) = -{8\pi G\o 3 c^2} {\trho\o g^{1/2}} + \mathscr{O}(\xi^2)
\ee

Now we are in a position to integrate (\ref{ksir}) for the scalar perturbation $\xi$ and (\ref{r-tt}) for the metric perturbation $\psi$. Since the sources in (\ref{5dee}) are constructed to be covariant in 5D \cite{wil2}, we must integrate the 5D density $\trho$ over the invariant 5D volume element $\int {\t g}^{1/2} dx^5 d^3x = \int g^{1/2} \phi dx^5 d^3x$. Therefore, a factor of $\phi$ is introduced to the integrals along with $g^{1/2}$. 

In practice, as is usual in a linear treatment, the metric and scalar potential are only carried to zeroth order in the source terms.\cite{sc} Therefore, in this calculation, we can define total rest mass $M$ in these coordinates simply as
\be
\label{intmass}
M \equiv \int \trho\ d^3x\ dx^5
\ee
We carry the $x^5$ integral formally, yet the cylinder condition, $\p_5{\t g}_{ab} =0$, insures that it amounts to just a constant factor that can be absorbed in $\t G$ in (\ref{5dee}). Its mathematical importance here is to include a factor of $\phi$ in the 5D covariant volume element.

Using the definition (\ref{intmass}), let us now integrate (\ref{ksir}) over all space and apply Stokes theorem to obtain
\be
\label{gradksi}
\left. {\p\xi\o\p r}\right|_{planet} = -{2\o 3}{GM\o r^2 c^2}
\ee
Here is the interesting result that the Kaluza scalar field perturbation is the same size as the metric perturbation. As Dicke \cite{dke}, \cite{dke2} noted, the strength of the scalar field is as {\it weak} as gravity, in that $\xi\ll 1$ like $\psi\ll 1$. For the purposes of a bouyancy concept, we realize that the scalar potential is as {\it large} as the gravitational potential. We shall see if the forces are as well. 

The magnitude of $\xi$ in (\ref{gradksi}) is consistent with the assumption (\ref{ksi}). It is perhaps counter-intuitive, that the dimensionless Newtonian gravitational potential $\psi_N \sim GM_\oplus/R_\oplus c^2$ at the surface of the earth is $\sim 10^{-10}$, a vanishingly small number. Yet its gradients create forces of engineering significance because of a coupling into mass energy $Mc^2$, which is a large number. We are contemplating something similar for the Kaluza scalar field. 

Equation (\ref{gradksi}) is a main result of this work, but is familiar from Brans \& Dicke (BD) \cite{bd} for $\omega=0$. The correspondence is because the BD theory has essentially the same equation for the scalar field, (\ref{trace}), based on their observation that the simplest covariant equation for the BD scalar field involves the D'Alembertian equated to the trace of the matter energy-momentum. 

The correspondence between BD theory, and neutral-matter 5-dimensional Kaluza general relativity, cannot be framed in terms of the BD free parameter $\omega$. The BD $\omega$ corresponds to a scalar field kinetic term in the Lagrangian, which is obviously not present in (\ref{Lag}). It arises in BD theory to generalize conformal transformations, which arise from transforming (\ref{Lag}), written in the ``Jordan frame", to a coordinate system in which the scalar field vanishes from the Ricci tensor term, the ``Einstein frame". Yet this transformation does not change the physics, because test particles still move on geodesics of the Jordan frame.\cite{sc2} Formally, the $\omega$ term seems only to add an awkward, {\it second} scalar-field energy-momentum tensor to the BD gravitational field equations.

More importantly, a 4D conformal transformation of the 5D metric (\ref{metric}) would impact the other terms in the Kaluza Lagrangian. This is because the identification of the 4D metric and electromagnetic potentials, (\ref{metric}), relies on the presence of a scalar field. Transforming the scalar field to a constant in (\ref{metric}) is akin to transforming to a frame in which the time-time component of the 4D metric is a constant. Put another way, the identification of the electromagnetic field in 5D is tantamount to the Jordan frame. We are free to work in the 5D Einstein frame, as in Ref.~\cite{coq}, but we find the Jordan frame more intuitive.

Now let us return to the gravitational field equations. The Kaluza scalar field acts to maintain the scalar curvature $R$ in (\ref{trace}) at zero, per (\ref{ksf}). Then we can write the gravitational field equation for $G_{tt}$ in (\ref{kee}), under the listed constraints:
\bea
\label{Gtt}
G_{tt} = R_{tt} &=& \nabla^2 \xi 
+ {8\pi G\o c^2}\trho + \mathscr{O}(\psi^2) \cr
&=& {16\pi G\o 3 c^2}\trho + \mathscr{O}(\psi^2) + \mathscr{O}(\xi^2)
\eea
where we used the scalar field equation (\ref{ksir}) in the last step, and where we are now carrying terms linear in both $\psi$ and $\xi$. 

The gravitational field equation (\ref{Gtt}) can now be readily integrated over all space, using (\ref{rtt}), (\ref{grtt}), and (\ref{intmass}), to obtain:
\be
\label{gradpsi}
\left. {\p\psi\o\p r} \right|_{planet}= {8\o 3}{GM\o r^2 c^2}
\ee

Let us now compare key features of the two potentials, $\psi$ (\ref{gradpsi}) and $\xi$ (\ref{gradksi}). One is that the magnitude of the scalar potential is $1/4$ the gravitational potential. Another is that the signs are opposite, with $\xi > 0$, while $\psi <0$, as usual. This means that the scalar potential associated with the mass $M$ is repulsive, and this is the origin of the electro-gravitic bouyancy concept derived here.

A third point is that the expression (\ref{gradpsi}) for the gravitational field in the presence of a scalar field differs from the usual Newtonian result, $\psi_N = 2GM/rc^2$. The difference between these two is that in the Kaluza picture, the effective gravitational mass-energy of a planet includes a contribution of mass-energy from the Kaluza scalar field bound to the mass. There is, therefore, a mass ``clothed" by the scalar field, and which accounts for Keplerian dynamics, and a ``bare" mass, which is the actual mass-energy devoid of scalar field mass-energy. Therefore we can make the identification
\be
\label{clth}
(GM)_{clothed} = {4\o 3} (GM)_{bare}
\ee
The Kaluza scalar field stores $1/3$ the mass-energy as the matter it is bound to, and the total gravitating mass-energy is increased above the bare mass by this amount. We keep the factor $G$ because a redefinition of $M$ in these terms is indistinguishable from a redefinition of $G$.

In his analysis of long range scalar fields \cite{dke}, \cite{dke2}, Dicke discussed how the scalar force would ``masquerade" as gravity. He specifically meant that the action of the scalar force would be indistinguishable from gravity. The Kaluza scalar field masquerades here as gravity, although in a different sense than Dicke meant. Here, in terms of the potentials, the action of scalar field mass-energy is indistinguishable from the action of material mass-energy, insofar as Keplerian dynamics is determined by the $g_{tt}$ component of the metric. 

The similarity to Dicke's conception of a long range scalar field breaks down when we consider the nature of matter coupling to that field.
%%%%%%%%%%%%%%%%%%%%%%%%%%%%%%%
%%%%%%%%%%%%%%%%%%%%%%%%%%%%%%%
\subsection{D. Scalar fields in the lab}
%%%%%%%%%%%%%%%%%%%%%%%%%%%%%%%
%%%%%%%%%%%%%%%%%%%%%%%%%%%%%%%%

Now we consider the Kaluza long-range scalar field around a body of mass $M$ and electric charge $Q$, anticipated to both be of magnitudes that are accessible in laboratory experiments. 

This corresponds to the case $\t U_5^2 \ll \xi^{-1}$. As in the neutral matter case, $cd\tau /ds \simeq 1$, and the mass (\ref{mass}) is just the rest mass $M$. There is now an electric charge $Q$, and a scalar charge $Mc^2 {\t U}_5^2$.

Let us start with the Maxwell equations, (\ref{max}), identifying $\trho ck {\t U}_5 \equiv \sigma$ as the electric charge density. 
\be
\label{maxweak}
g^{-1/2}\p_r \left(g^{1/2}\phi^3 {E^r/ c}\right) = \mu_0 c \sigma
\ee
where $E^r$ is the radial component of the electric field. Recall that $\phi^3 g^{1/2} = r^2 \sin\theta + \mathscr{O}(\psi) + \mathscr{O}(\xi)$. For the electric field equation, we will write explicit terms to zeroth order in the perturbations, but still keep track of the ordering. 

Let us now introduce the total charge integral $Q$, analogous to the mass integral in (\ref{intmass}):
\be
\label{intchg}
Q \equiv \int \trho ck {\t U}_5\ d^3x\ dx^5
\ee
Let us use this definition of electric charge to integrate (\ref{maxweak}) over all space, and use Stokes theorem to obtain
\be
\label{E}
E^r = {1\o 4\pi\epsilon_0} {Q\o r^2} + \mathscr{O}(\psi) + \mathscr{O}(\xi)
\ee

This is clearly the typical Coulomb expression. The perturbation in $\psi$ is a normal Newtonian perturbation to the relativistic Maxwell equations. 

A perturbation effect from the Kaluza scalar field would be new. Yet because of the presumed small size of $\xi$, its effects are likely to be minor, and masked as an effective dielectric constant, as in the Bekenstein theory of scalar-electromagnetic coupling \cite{bk}. In practice, they might be difficult to distinguish also from curvature effects in $\psi$, which are of the same magnitude.

Let us now consider the Kaluza scalar field perturbation described by (\ref{ksffull}).
%\begin{comment}
\bea
\label{ksfp}
{3\o g^{1/2}}&\p_r& \left(g^{1/2} \p^r \xi \right) =
\mu_0 k^2 c^2 {\trho \o g^{1/2}} {{\t U}_5^2\o\phi^2}\cr
&-& {8\pi G\o c^2}{\t\rho \o g^{1/2}} 
+ {3\o 2} k^2 \phi^3 E^2/c^2
+ \mathscr{O}(\xi^2)
\eea
where we used $F^{\a\b}F_{\a\b}=-2E^2/c^2$.
%\end{comment}

We can use the zeroth order expression for the electric field from (\ref{E}) in (\ref{ksfp}). We see that it is essentially the usual electrostatic mass, $M_E$, defined as the integral of the electrostatic energy density over all space:
\be
\label{esmass}
M_E c^2 \equiv \int {1\o 2}\epsilon_0 E^2 d^3 x
\ee

Now let us define the scalar charge integral $S$, as we did for mass (\ref{intmass}) and electric charge (\ref{intchg}).
\be
\label{intschg}
S \equiv c^2 \int \trho {\t U}_5^2 \ d^3x\ dx^5 
\ee
Such an expression for scalar charge is unique and has no analog in conventional physics. Its units are energy. It can be understood as the integral of the specific charge squared per unit rest mass of the source. For a uniform source, we can write
\be
\label{simpintschg}
S = Q^2/Mk^2
\ee
consistent with (\ref{schg}). We also see that this implies the coupling coefficient for scalar charge in (\ref{ksfp}) is $\mu_0 k^2 = 16\pi G/c^4$.

Now we can use these expressions to integrate the Kaluza scalar field equation (\ref{ksfp}) over all space, to find:
\be
\label{ksfqm}
{\p\xi\o\p r} = {\mu_0\o 12\pi}{Q^2/M\o r^2} - {2G\o 3c^2} {M\o r^2} + {4G\o c^2}{M_E\o r^2}
+ \mathscr{O}(\xi^2)
\ee

The expression (\ref{ksfqm}) for the scalar field perturbation has some interesting features. One is that neutral matter sources have the opposite sign as charged matter and as electric fields. That is, electric charge and electric fields are attractive sources of Kaluza scalar field, while neutral matter is a repulsive source of the scalar field, much like positive and negative electric charges can act as attractive or repulsive sources of electric field.

Here also is a key result of this work, that a third electro-gravitic lengthscale, $\mu_0 Q^2/M$, appears alongside the two lengthscales that characterize the Reissner-Nordstrom metric, $GM/c^2$ and $Q\sqrt{G/\epsilon_0 c^4}$. A third electro-gravitic lengthscale would seem to implicate new physics not anticipated in electrodynamics and general relativity.

The term in $Q^2$ will dominate laboratory scalar fields. A typical laboratory capacitance might be $10^{-10}$ farad, and typical lab voltages are $10^3$ volts. Therefore, laboratory charges $Q_{lab}\sim 10^{-7}$ coulombs. For sizes of order 1 meter and masses of order 1 kilogram, the potential for the term in the scalar charge is of order $10^{-22}$. This is much smaller than the planetary scalar potential, of order $10^{-10}$, yet much larger than the mass term in (\ref{ksfqm}), of order $10^{-27}$. The term in the electrostatic mass $M_E$ is even smaller. Therefore we can approximate the laboratory Kaluza long-range scalar field of a charged, massive object:
\be
\label{ksflab}
\left. {\p\xi\o\p r} \right|_{lab} \simeq {\mu_0\o 12\pi}{Q^2/M\o r^2} 
\ee
The rough order of magnitude of $\xi_{lab} \sim 10^{-22}$, which is fully 10 orders of magnitude smaller than the Kaluza scalar field of the earth calculated above.

The scalar potential (\ref{ksflab}) can be compared with the weak-field solution of Chodos \& Detweiler \cite{cd}. It is similar quantitatively to findings reported here, but is opposite sign because those authors choose the 5th coordinate signature to be spacelike. In that case, the scalar field behaves like gravity, and is attractive. That is also true of Ferrari's sign choice.

Let us now consider the gravitational field equation (\ref{kee}), specifically the $R_{tt}$ component. It has sources in matter, electromagnetic field, and scalar field mass-energy. The action of the scalar field will act to maintain the scalar curvature $R$ according to (\ref{ksf}):
\be
\label{Ru}
R = -\mu_0 k^2 c^2 {\trho\o g^{1/2}\phi}{{\t U}_5^2\o\phi^2} - {3\o 2} k^2 \phi^2 E^2/c^2
\ee

The $tt$ component of the scalar field energy-momentum tensor is
\bea
\label{Ttt}
T_{tt}^\phi &=& \nabla^2 \xi + \mathscr{O}(\xi^2) \cr
&=& {\mu_0\o 3} k^2 c^2 {\trho \o g^{1/2}} {{\t U}_5^2\o\phi^2}
- {8\pi G\o 3 c^2}{\t\rho \o g^{1/2}} \cr
&+& {1\o 2} k^2 \phi^3 E^2/c^2 + \mathscr{O}(\xi^2)
\eea
where the second equality follows from (\ref{ksfp}).

The term in (\ref{kee}) in the $tt$ component of the electromagnetic energy-momentum tensor is
\be
\label{eme}
{8\pi G\o \mu_0 c^4}\phi^2 T^{EM}_{tt} = {8\pi G\o c^4}\phi^2 {1\o 2}\epsilon_0 E^2
\ee

The field equation for $\psi$ is obtained when we combine (\ref{kee}), (\ref{Ru}), (\ref{Ttt}), (\ref{eme}), (\ref{r-tt}), and (\ref{grtt}):
\be
\label{psidiff}
{1\o g^{1/2}}{\p_r} \left( g^{1/2} \p_r \psi /2 \right) =
{\trho \o g^{1/2}\phi } \left( {16\pi G\o 3c^2} - {\mu_0\o 6} k^2 c^2 {{\t U}_5^2\o \phi^2}\right)
\ee
The terms in the electric field cancel.

Now integrate (\ref{psidiff}) over all space and use Stokes theorem with (\ref{intmass}) and (\ref{intschg}) to obtain
\bea
\label{gradpsilab}
\left. {\p\psi\o\p r} \right|_{lab} &=& {8\o 3}{GM\o r^2 c^2} - {\mu_0\o 12\pi}  {S\o r^2}\cr
&\simeq & - {\mu_0\o 12\pi}  {S\o r^2}
\eea

The expression in (\ref{gradpsilab}) is seen in the neutral matter case, (\ref{gradpsi}), but here the $M$ is understood to be of laboratory dimensions. As discussed for (\ref{ksfqm}), the term in $M$ is of order $10^{-27}$ and negligible in the laboratory, as we expect for laboratory gravitational effects. Since the term in $S$ is larger, it seems to imply the scalar charge can swamp the matter charge in the sourcing of the gravitational field. Yet these considerations are for laboratory scale only. It still appears no laboratory charge could counteract the Kaluza scalar field of a planet.

The scalar charge acts negatively as a source of gravitational field in (\ref{gradpsilab}). The terms in $M$ and $S$ are also in the expression for $\xi$, (\ref{ksfqm}). In that expression, matter creates a repulsive potential as we saw previously, and scalar charge an attractive potential. Therefore, scalar charge and mass act oppositely as sources of the gravitational and Kaluza scalar fields. A scalar charge source creates equal and opposite perturbations of the metric and of the scalar field, as seen by comparing (\ref{gradpsilab}) and (\ref{ksflab}).

Note that (\ref{ksfqm}) and (\ref{gradpsilab}) sum to:
\be
{\p\psi\o\p r} + {\p\xi\o\p r} \simeq {2GM\o r^2 c^2} +
 \mathscr{O}(\xi^2) + \mathscr{O}(\psi^2)
\ee
which is the ordinary Newtonian potential in terms of the bare mass.
%%%%%%%%%%%%%%%%%%%%%%%%%%%%%%%
%%%%%%%%%%%%%%%%%%%%%%%%%%%%%%%
\subsection{E. Atomic scalar fields}
%%%%%%%%%%%%%%%%%%%%%%%%%%%%%%%%%%%%%%%%%%%%%
%%%%%%%%%%%%%%%%%%%%%%%%%%%%%%%%%%%%%%%%%%
We have seen that planetary $\xi \le 10^{-10}$, consistent with the perturbation expansion (\ref{ksi}). Laboratory-generated $\xi \sim 10^{-22}$, much smaller still. Achievable laboratory values of ${\t U}_5 \sim 10^3$, as defined in (\ref{echg}). The terms in $\xi {\t U}^2_5$ in (\ref{mass}) and (\ref{schg}) are therefore negligible for all values of ${\t U}_5 < 10^{5}$. This means that the saturated limits (\ref{massx}) and (\ref{schgx}) of (\ref{mass}) and (\ref{schg}) are achieved only for very high charge-to-mass ratios. Such ratios are only found in atomic systems. The electron ${\t U}_5 \vert_e \sim 10^{21}$ and the proton ${\t U}_5 \vert_p \sim 10^{18}$. It appears the saturated limits are appropriate for atomic systems.

Following Rohrlich, we direct the classical theory at elementary charged particles only insofar as we can make predictions that are independent of particle structure, since that is outside the domain of validity of the classical theory. Yet the classical theory can be used to address atomic systems in a structure-independent way. In practical terms, it means assuming a structure of characteristic size $r_0$, and then letting $r_0 \rightarrow 0$. If the resultant quantity is finite and well-behaved, it is a valid calculation. If the result is infinite, then it indicates a failure in the application of the theory. One example of such failure, as pointed out by Rohrlich, is the model of a point particle. If the classical model is a charged sphere, then the electric field energy goes to infinity as the sphere size goes to zero, and therefore the point particle is not a valid classical model of a charged particle.

In the limit that ${\t U}_5^2 \gg \xi^{-1}$, mass (\ref{mass}) $\rightarrow M{\t U}_5\sqrt{2\xi}$ (\ref{massx}), and scalar charge (\ref{schg}) $\rightarrow Mc^2 {\t U}_5/\sqrt{2\xi}$, (\ref{schgx}). Now there is a formal convergence in that the mass, electric charge, and scalar charge, are all proportional to ${\t U}_5$. The 3 forces masquerade as the Coulomb electric force at ultra-high specific charge states. We will calculate $\xi$ and then double check that our assumptions are satisfied, and the saturation limit is correct.

The Coulomb electric force in this limit is the same as the previous case, given by (\ref{E}) to zeroth order in the perturbations. There is no saturation effect present in the Maxwell equations (\ref{max}) or in the electric charge (\ref{echg}).

Consider now the Kaluza long-range scalar field equation (\ref{ksffull}) once more, similar to (\ref{ksfp}), but with saturated charges:
\bea
\label{ksfep}
{3\o g^{1/2}}&\p_r& \left(g^{1/2} \p^r \xi \right) =
\mu_0 k^2 c^2 {\trho \o g^{1/2}} {{\t U}_5^2\o\sqrt{2\xi}}\cr
&-& {8\pi G\o c^2}{\t\rho \o g^{1/2}}\sqrt{2\xi} 
+ {3\o 2} k^2 \phi^3 E^2/c^2
+ \mathscr{O}(\xi^2)
\eea
The second term on the RHS is of first order in $\xi$ relative to the first term, and so can be ignored, consistent with our approximation of the sources of perturbations to only zeroth order in those perturbations. Such ordering is standard in the Newtonian limit of general relativity.\cite{sc}

Now let us integrate (\ref{ksfep}) from a minimum radius $r_0$, which we nominally take to be $10^{-10}$ meters. The first term on the RHS of (\ref{ksfep}) has a dependence on $\xi$, which we take to be the value of $\xi$ evaluated at the particle. We assume no structure inside $r_0$, and so take $\xi_0$ to be the constant value of $\xi$ inside $r_0$. It therefore forms a boundary value on the source term in the integral. Using (\ref{intchg}) and (\ref{E}) we find:

\bea
\label{ksfepi}
{\p\xi\o\p r} &\simeq& {\mu_0\o 12\pi}{kc\o\sqrt{2\xi_0}}{Q\o r^2} +  {\mu_0\o 2\pi}{G\o c^2 r_0} {Q^2\o r^2} \cr
&\simeq& {\mu_0\o 12\pi}{kc\o\sqrt{2\xi_0}}{Q\o r^2}
\eea
We find that the term in $Q^2$ is of order $10^{-20}$ smaller than the term linear in $Q$ at $r_0 \sim 10^{-10}$ m, and so we ignore it. Eventually the Coulomb term diverges as $r_0$ goes to zero, but that is not particular to the Kaluza theory.

Now solve for $\xi_0$ by evaluating (\ref{ksfepi}) at $r_0$ to find
\be
\label{ksi0}
\xi_0 = \left({ \mu_0 Q kc \o 12\sqrt{2}\pi r_0 }\right)^{2/3}
\ee
For typical atomic paramters, $\xi_0 \sim 10^{-17}$, still $\ll 1$ in the high charge states of atomic systems. Therefore the atomic value of the Kaluza scalar field from an elementary particle is given by
\be
\label{ksfatom}
\left.{\p\xi\o\p r} \right\vert_{atomic} \simeq \left({ \mu_0 Q kc \o 12\sqrt{2}\pi}\right)^{2/3} {r_0^{1/3}\o r^2}
\ee
Note this scalar potential behaves gravitationally in that like charges attract, but it appears opposite charges repel, even as it is apparently proportional to electric charge in this limit. We will compare to the atomic Coulomb force shortly. 

It appears from this analysis that, as $r_0 \rightarrow 0$, $\xi_{atomic} \rightarrow 0$. Therefore, the theory seems to imply a structure-independent stability of the Kaluza scalar field for point particles. This is because the scalar field enters in the denominator of the source term in (\ref{ksfep}). As the scalar field strength increases, it reduces the magnitude of its own source.

Now let us consider the gravitational field equation for $\psi$, and the saturated mass (\ref{massx}). We consider the time-time component of the gravitational field equations (\ref{kee}), with $R$ obeying the Kaluza scalar field equation (\ref{ksf}), the time-time component of the scalar field energy-momentum tensor (\ref{tphi}) given by (\ref{ksffull}), and the time-time component of the electromagnetic energy-momentum given by (\ref{eme}), to obtain an equation very similar to the low-charge equation for $\psi$ (\ref{psidiff}):
\bea
\label{psix}
{1\o g^{1/2}}&{\p_r}& \left( g^{1/2} \p_r \psi /2 \right) \cr &=&
{\trho \o g^{1/2}\phi } \left( {16\pi G\o 3c^2}{\t U}_5 \sqrt{2\xi} - {\mu_0\o 6} {k^2 c^2\o  \sqrt{2\xi}} {{\t U}_5\o \phi^2}\right)
\eea
where the electric terms again have cancelled. The term in $G$ from the usual matter source is order $\xi$ smaller than the term in $\mu_0$, the scalar source term in $\psi$. Therefore we can drop the term in $G$ to this approximation, and keep only the term in $\mu_0$. Then the gravitational potential at atomic scale is given by
\be
\label{gradpsiatm}
\left. {\p\psi\o\p r} \right|_{atomic} 
\simeq  - {\mu_0\o 12\pi}  {Q\o r^2}{kc\o\sqrt{2\xi_0}}
\ee
Once more, we have set $\xi_0$ to be the value of $\xi$ at the boundary of the source $r_0$, and then a limit is taken as $r_0 \rightarrow 0$. We see that the gravitational field is actually repulsive in this limit, and opposite to the scalar potential, which is attractive. However, the scaling of the two potentials $\xi$ and $\psi$ is the same in the saturated limit, and both obey (\ref{ksfatom}). It appears the saturated gravitational potential goes to zero as $r_0\rightarrow 0$, along with the saturated scalar potential.

Having completed an evaluation of the scalar, electric, and gravitational fields for charged and neutral sources, let us turn to the implications for those fields on the motion of test bodies.

%%%%%%%%%%%%%%%%%%%%%%%%%%%%%%%%%%%%%%%%%%%%%
%%%%%%%%%%%%%%%%%%%%%%%%%%%%%%%%%%%%%%%%%%
\section{IV. SCALAR, ELECTRIC, AND GRAVITATIONAL FORCES}
In this section, we combine the previous expressions for charges and potentials to obtain the forces between massive, charged objects. As for the potentials, we consider the fields of neutral, weakly-charged, and strongly-charged systems. A unique electro-gravitic buoyant lift force is discovered.

\subsection{A. Lift in the planetary scalar field}
%%%%%%%%%%%%%%%%%%%%%%%%%%%%%%%%%%%%%%%%%%%%%
%%%%%%%%%%%%%%%%%%%%%%%%%%%%%%%%%%%%%%%%%%
Consider now the motion of an electrically-charged test particle of mass $m$ and charge $Q$, where ${\t U_5}^2\ll \xi^{-1}$, obeying the equation of motion (\ref{4Deom}), moving in the planetary, neutral-matter potentials described in (\ref{gradksi}) and (\ref{gradpsi}). The particle will of course couple to the gravitational field irrespective of its charge or mass, according to the equivalence principle, but there is no ambient electric field to couple with. The scalar charge (\ref{schg}) is given by (\ref{simpintschg}), but now with the rest mass as $m$. The mass (\ref{mass}) is just $m$.

Consider then the radial component of (\ref{4Deom}):  
\bea
m{dU^r\o dt} &+& {4\o 3}{GMm\o r^2} = {(Q/m)^2\o 16\pi G\epsilon_0}
{2\o 3}{GMm\o r^2} \cr \noalign{\vskip7pt}
&+& \mathscr{O}(U^2/c^2) + \mathscr{O}(\psi^2) + \mathscr{O}(\xi^2)
\eea
The quantity $M$ in the preceding analysis is understood to be the bare mass of the planet. Let us convert to the clothed mass $M_{cl}$ from (\ref{clth}), corresponding to the mass measured in gravitational experiments, and now ignore the second-order terms:
\be
\label{egl}
m{dU^r\o dt} + {GmM_{cl}\o r^2} \simeq {(Q/m)^2\o 16\pi G\epsilon_0}
{GmM_{cl}\o 2r^2}
\ee

The scalar force in (\ref{egl}) acts counter to Newtonian gravity, and in proportion to the gravitational weight of the test body. Yet its coupling is quasi-electric. Therefore, we consider this an ``electro-gravitic" bouyancy effect. Against the gravitational weight, the electric charge appears as an enhanced effective weight. In fact, we can define an electro-gravitic mass, whose upward force is equal to the weight of the mass:
\be
\label{meg}
m_{EG} = {Q^2/m\o 32\pi G\epsilon_0}
\ee

The electro-gravitic constant (\ref{k}) sets a charge-to-mass scale of $10^{-10}$ C/kg. In laboratory settings, charges are conveniently expressed in terms of capacitance $C$ and electric potential $V$, $Q=CV$. Typical capacitances for meter-sized objects are $10^{-10}$ farad. An object of mass $10$ kg, charged to $1000$ V, would feel a scalar force of $10^4$ times its weight.

This scalar force is quite different than the one contemplated by Dicke, because the Kaluza scalar coupling is electrostatic. The field equations for the scalar field are the same, but the scalar charge is different. Indeed, the Brans-Dicke scalar field is explicitly barred from the force equations, and enters the field equations only. Therefore, the Kaluza scalar force would not masquerade as gravity, as Dicke anticipated. Dicke viewed the scalar field as attractive to mass like gravity, and coupling to mass like gravity. The electric coupling of the Kaluza theory leads to vastly larger scalar forces than Dicke anticipated based on his field equations, and opposite in direction.

%%%%%%%%%%%%%%%%%%%%%%%%%%%%%%%%%%%%%%%%%%%%%
%%%%%%%%%%%%%%%%%%%%%%%%%%%%%%%%%%%%%%%%%%
\subsection{B. Forces between neutral masses}
%%%%%%%%%%%%%%%%%%%%%%%%%%%%%%%%%%%%%%%%%%%%%
%%%%%%%%%%%%%%%%%%%%%%%%%%%%%%%%%%%%%%%%%%
Now let us consider the forces between 2 neutral massive bodies of bare rest mass $M_1$ and $M_2$, separated by a distance $r$. Their electric charge is zero, and so there is no electric interaction. However, there will be both gravitational and scalar forces.

Except for the effect of the clothed mass discussed above, the gravitational force between massive bodies abides the conventional Newtonian limit. The gravitational potential $\psi_M$ of a bare mass $M$ is given by (\ref{gradpsi})
\be
\psi_M = -{8\o 3}{GM\o rc^2} = -  {2GM_{cl}\o rc^2}
\ee
The mass (\ref{mass}) is just the bare rest mass $M$. Therefore the gravitational force between the two bodies is
\be
F_g = {4\o 3}{GM_1 M_2\o r^2} = {GM_{1cl}M_2\o r^2}
\ee
This illustrates that the potential depends on the clothed mass, but the coupling depends on the bare mass.

The scalar potential of bare mass $M$ is given by (\ref{gradksi}):
\be
\xi_M = {2\o 3}{GM\o r c^2}
\ee
In this case, the scalar charge (\ref{schg}) is zero, because $\U5 =0$. Therefore, the scalar force between neutral masses is zero. 

The Kaluza scalar field contributes indirectly to the gravitational interaction through its own energy density, but no scalar force manifests between neutral masses. This is similar to the Brans-Dicke theory, except they posit from the outset that the scalar field can have no effect on the motion of material bodies. 

%%%%%%%%%%%%%%%%%%%%%%%%%%%%%%%%%%%%%%%%%%%%%
%%%%%%%%%%%%%%%%%%%%%%%%%%%%%%%%%%%%%%%%%%
\subsection{C. Forces between weak charges}
%%%%%%%%%%%%%%%%%%%%%%%%%%%%%%%%%%%%%%%%%%%%%
%%%%%%%%%%%%%%%%%%%%%%%%%%%%%%%%%%%%%%%%%%
Consider now two charges, $Q_1$ and $Q_2$, of mass $M_1$ and $M_2$, and such that ${\t U}_5^2 \ll \xi^{-1}$ for both. They will experience mutual gravitational, electric, and scalar forces.

The scalar potential generated by $Q_1$ is given by (\ref{ksflab}):
\be
\label{ksflab1}
\left. {\p\xi\o\p r} \right|_{1} \simeq {\mu_0\o 12\pi}{Q_1^2/M_1\o r^2}
\ee
Meanwhile, the scalar charge of $Q_2$ is given by (\ref{simpintschg}), $Q_2^2/M_2 k^2$. Therefore, the scalar force between $Q_1$ and $Q_2$ is:
\be
F^r_S = {-1\o 16\pi G \epsilon_0^2} {Q_1^2\o M_1}{Q_2^2\o M_2}{1\o r^2}
\ee
This means that the scalar force is attractive between charges, irrespective of sign. In this way, it is like gravity. Let us compare it to the usual Coulomb force:
\be
F^r_E={1\o 4\pi\epsilon_0}{Q_1 Q_2\o r^2}
\ee

The ratio of the two forces is seen to be:
\be
\label{frat}
{F^r_S\o F^r_E} = {1\o 3}{(Q_1/M_1)(Q_2/M_2)\o 16\pi G\epsilon_0}
\ee
For achievable laboratory charge-to-mass ratios of order $10^{-7}$ C/kg, the ratio of forces can be of order $10^5$, extremely large. Here is a verifiable difference from Coulomb's law, in that the additional scalar force can be large in laboratory environments. 

Ferrari \cite{fri} also examined deviations from Coulomb's law in this theory. He considered time-independent scalar and electric forces between charged objects, and found a potentially significant variation from the Lorentz force law, as we are finding in (\ref{frat}). However, Ferrari made significantly different assumptions than used here, and his results may have been compromised by them. 

Ferrari obtained his solution by assuming all fields could be expanded in powers of the two lengthscales he identified in the system: $GM/c^2$ and $Q\sqrt{G/4\pi\epsilon_0 c^4}$. Yet the solution seen in (\ref{ksflab}) indicates there is a third lengthscale, $\mu_0 Q^2 / M$. Ferrari also seems to miss the scalar field contribution to gravitational mass seen in (\ref{clth}). Finally, Ferrari notes some discrepancies between his results and conventional limits. Therefore, while the general prediction of deviations from the Coulomb force between charges due to a scalar interaction was predicted by Ferrari, the magnitudes and mathematical scaling of these results appears to be different. 

The general result of scalar-induced deviations from Coulomb's law seen by various researchers seems to invite an experimental investigation. The experimentalist should note this is not a deviation in the $1/r^2$ geometric part of Coulomb's law, but a deviation in the magnitude of the force.

%%%%%%%%%%%%%%%%%%%%%%%%%%%%%%%%%%%%%%%%%%%%%
%%%%%%%%%%%%%%%%%%%%%%%%%%%%%%%%%%%%%%%%%%
\subsection{D. Forces between strong charges}
%%%%%%%%%%%%%%%%%%%%%%%%%%%%%%%%%%%%%%%%%%%%%
%%%%%%%%%%%%%%%%%%%%%%%%%%%%%%%%%%%%%%%%%%
We previously recognized that only atomic particles satisfy the requirement for strong charges, $\U5 \gg \xi^{-1}$. Let us then consider bodies of charge $Q_1$ and $Q_2$, and masses $M_1$ and $M_2$.

For atomic sources of mass and electric charge, the electric field and electric force have their usual form as in (\ref{E}), and no modifications need be discussed.

The saturated gravitational (\ref{gradpsiatm}) and scalar (\ref{ksfepi}) potentials are equal and opposite. The saturated gravitational charge is a factor $\xi$ smaller than the saturated scalar charge. Therefore the attraction between charges is approximately the scalar attraction: 

\be
F^r_{S+} \propto {Q_1 Q_2\o \epsilon_0 r^2}{1\o\sqrt{\xi_1 \xi_2}}
\ee

In the saturated regime characteristic of atomic systems, the scalar force masquerades as the electric force, similar to the way Dicke anticipated it would masquerade as the gravitational force. Yet as shown in (\ref{ksfatom}), the scalar potential would appear to go to zero as the source goes to a point particle, so that the scalar interaction is strongly suppressed for atomic systems.

Yet as we pass from outer atomic scales of $10^{-10}$ m, and approach point particles, the electrostatic energy must be considered. Since the Kaluza scalar field is attractive for like charges, akin to gravity, then it might provide a stabilizing influence in the energy budget of the point particle, as shown already by Ref.~\cite{adm} for Coulomb electric fields.

For the purposes of a classical theory, $F^r_{S+} \rightarrow 0$ as $r_0 \rightarrow 0$.The vanishing of the scalar force at atomic scales is accompanied by a masquerading of the electric force in that regime, in that the scalar charge could become linear in electric charge at certain high specific charge states.

%%%%%%%%%%%%%%%%%%%%%%%%%%%%%%%%%%%%%%%%%%%%%
%%%%%%%%%%%%%%%%%%%%%%%%%%%%%%%%%%%%%%%%%%
\subsection{E. Forces between charged and neutral bodies}
%%%%%%%%%%%%%%%%%%%%%%%%%%%%%%%%%%%%%%%%%%%%%
%%%%%%%%%%%%%%%%%%%%%%%%%%%%%%%%%%%%%%%%%%
Consider now a body of charge $Q$ and mass $M_Q$, and a body of mass $M$ and zero charge. We consider non-atomic systems, so that ${\t U}_5\ll \xi^{-1}$.

The electric charge identified in (\ref{echg}) and (\ref{U5}) implies an induced electric charge for any neutral body of mass $M$ at rest in an electric field:
\be
{\t U}_5 = \phi^2{d\tau\o ds}kA_0 U^0 \simeq \sqrt{G\o\pi \epsilon_0}{Q\o rc^2} + \mathscr{O}(\xi) 
\ee
This implies the resulting induced electric charge, which depends on $M$:
\be
Q_M = Q {4GM\o rc^2} \sim Q  \psi \ll Q
\ee
This means the electric field generated by the charge $Q$ will induce an electric charge in the neutral mass $M$. It is clear from the appearance of $G$ that this can be considered an electro-gravitic effect. But since the metric perturbation $\psi\ll 1$, the induced charge is likewise small. Nonetheless, there would be an induced electrostatic repulsion of order $\psi$. Under laboratory conditions, $\psi \sim 10^{-28}$, so the effect is probably not observable in the laboratory.

The induced electric charge also implies an induced scalar charge, as given by (\ref{schg}) and (\ref{simpintschg}):
\be
S_M = {Q_M^2\o Mk^2} = \left( Q^2\o \pi\epsilon_0 r \right) \left(MG\o rc^2 \right) \sim M_E c^2 \psi
\ee
using the definition (\ref{esmass}) for electrostatic mass. As with the induced electric charge, the induced scalar charge is the electrostatic potential energy of the charge Q at M, modulated by $\psi$, so the induced scalar charge is a perturbation of order $\psi$.

With $S_M$ for the induced scalar charge, and (\ref{ksflab}) for the scalar potential generated by the charge, we can write the induced scalar force between a neutral body of mass $M$ and a body of electrical charge $Q$:
\be
F_{Si} \sim {GM_E M\o r}{\mu_0\o 12\pi}{Q^2/M\o r^2} \propto G\mu_0 {Q^4\o r^4}
\ee
This force appears to be of order $10^{-50}$ N in the lab, too small to be detectable.

Now consider the reverse force, that due to the scalar force exerted by the scalar field of the neutral mass $M$ on the charge $Q$. The scalar charge is given by (\ref{schg}) and (\ref{simpintschg}):
\be
S_Q = {Q^2\o M_Q} ({c^2/ 16\pi G\epsilon_0})
\ee
The scalar field of the mass $M$ is given by (\ref{gradksi}), so that the scalar force exerted by $M$ on the charge $Q$ is given by:
\be
F^r_{SQ} = - {M\o M_Q}{1\o 24\pi\epsilon_0} {Q^2\o r^2}
\ee
This scalar force exerted by the mass $M$ on the charge $Q$ is much larger than the induced scalar force exerted by the charge $Q$ on the mass $M$. It results in a net attraction between charged and neutral bodies, at a strength similar to the strength of the Coulomb force of two bodies of equal charge. The magnitude seems too large to have gone undetected, but there can be an attraction between charged and neutral bodies as the electric field of the charged body polarizes the neutral one.

%%%%%%%%%%%%%%%%%%%%%%%%%%%%%%%%%%%%%%%%%%%%%
%%%%%%%%%%%%%%%%%%%%%%%%%%%%%%%%%%%%%%%%%%
\section{V. TUNING THE SCALAR COUPLING TO ZERO}
%%%%%%%%%%%%%%%%%%%%%%%%%%%%%%%%%%%%%%%%%%%%%
%%%%%%%%%%%%%%%%%%%%%%%%%%%%%%%%%%%%%%%%%%

We have in several cases encountered forces from the Kaluza scalar field that are apparently large, according to the mathematics, yet the field equations otherwise reproduces 4D physics. It seems unlikely that effects of the magnitude described here will be validated in the laboratory, for they should have already been discovered by now, if they exist. The reason is the possibility of large relative size of the scalar charge relative to the gravitational and electromagnetic charge in the force equation, (\ref{4Deom}).

We might therefore ask whether we can tune the theory to zero out the scalar interaction. We have seen that the atomic scalar interaction appears to vanish by virtue of the preceding analysis, leaving only macroscopic and laboratory-scale effects. To zero out the scalar force, either the field must be zero, or the coupling charge must be zero.

It is very difficult to hide the effects of the Kaluza long-range scalar field in the field equations. A conformal transformation to the Einstein frame would eliminate explicit force terms from the equations of motion, but particle geodesics in the Einstein frame still reflect the scalar influence, compared to its absence in the Jordan frame. Also, the early arguments by Dicke, and the results here, show that the Kaluza scalar field can masquerade as gravity in the absence of other couplings. Therefore we consider it more likely that the scalar charge is somehow zeroed out in a way our analysis has not grasped so far.

There is only one free parameter in the 5D field equations (\ref{5dee}), and that is the invariant length element $\t a$ of the source 5-velocity. The fundamental relation is given by (\ref{a2u}), and we assigned the value $\t a^2 = 1 + {\t U}_5^2$ as the only natural choice, given that 4D physics is recovered in the limit that $\phi \rightarrow 1$, and given that we desire the scalar mass function (\ref{mass}) $cd\tau /ds$ to be positive. 

Chodos \& Detweiler \cite{cd} also discuss tuning the scalar interaction to zero by an appropriate choice of $\t a^2$, so that the difference between the matter terms in (\ref{ksffull}) goes to zero. They find no good explanation for why that should be so, nor do we. But if that does account for the observed absence of the scalar interaction, the Kaluza scalar field should still clothe planetary masses.

Instead of choosing (\ref{a2c}), we might instead set $\t a^2$ in (\ref{a2u}) such that ${\t U}_5^2/\phi^2 \sim 0$ in (\ref{a2u}). But this is tantamount to setting ${\t U}_5=0$, if a 4D limit is to be obtained in (\ref{Lag}) as $\phi \rightarrow 1$. And then correspondence with the Lorentz force law is lost. In other words, it is difficult to tune $\t a$ to make the scalar charge (\ref{3scalar}) go to zero without also driving the electric charge (\ref{3charge}) to zero. If such large force effects are not validated, this may be the first testable classical falsification of the 5D hypothesis, that electromagnetism and gravity are aspects of a 5D metric.

%%%%%%%%%%%%%%%%%%%%%%%%%%%%%%%%%%%%%%%%%%%%%
%%%%%%%%%%%%%%%%%%%%%%%%%%%%%%%%%%%%%%%%%%
\section{VI. SCALAR FIELD LENGTHSCALES}
%%%%%%%%%%%%%%%%%%%%%%%%%%%%%%%%%%%%%%%%%%%%%
%%%%%%%%%%%%%%%%%%%%%%%%%%%%%%%%%%%%%%%%%%
Early in development of the Kaluza scalar field theory, it was realized that Kaluza's original assumption that $\phi\rightarrow 1$ was incompatible with the scalar field equation (\ref{ksffull}). When sources $\t\rho\rightarrow 0$, an unnatural constraint is implied for the electromagnetic field, $F^{\a\b}F_{\a\b}=0$. Conversely, the existence of ambient electromagnetic fields will influence $\phi$, driving it away from $1$. Yet we may ask how $\phi \rightarrow 1$ in a self-consistent fashion. 

Our analysis of $\phi$ must keep in mind the hierarchy of lengthscales at issue. It is similar in this respect to gravity. On laboratory lengthscales, and locally in any gravitational field, the local gravitational field is the Minkowski metric. On cosmological lengthscales, the gravitational field is the Robertson-Walker metric; and on non-cosmological timescales, the Robertson-Walker metric of the universe approximates the Minkowski metric, so that the spacetime of the universe is flat on time slices. Yet around massive objects, we encounter other gravitational fields, such as the Schwarzschild metric or the Kerr metric. 

All of these gravitational fields exist simultaneously, and they overlap in space and time. How are they distinguished? By the lengthscale under consideration. Just as the gravitational field can be viewed as either strongly curved or flat, depending on the lengthscale under consideration, so it is with the Kaluza scalar field. That is, regions of local scalar field variation where $F^{\a\b}F_{\a\b}\ne 0$ can coexist with scalar fields on different lengthscales, for which $F^{\a\b}F_{\a\b}=0$.

Indeed, this is the case when we consider scalar field cosmology. We recall that long range scalar field research was founded on study of a variable gravitational constant, which is cosmological by definition. Therefore, if $\phi \rightarrow 1$ cosmologically, if it is to be identified with the gravitational constant in the Lagrangian (\ref{Lag}), then the value of $F_{\mu\nu}$ in that limit must also be cosmological.

We find that the cosmological equation for the Kaluza scalar field $\phi_c$, when incorporating the known cosmological mass-energy density, and setting the scalar field to a constant, results in
\be
\phi_c^3 = {\mu_0 \rho_c c^2\o 3B_c^2}
\ee
where $\rho_c c^2$ is the cosmological energy density, and $B_c$ is a cosmological magnetic field. Clearly, this involves the ratio of matter energy density to bulk cosmological magnetic energy density. The cubic dependence of $\phi_c$ on those parameters results in a very weak dependence, and stability of $\phi$ around $1$ for a range of parameters.

The scalar field equation does allow a consistent, constant scalar field solution that can be identified cosmologically with the gravitational constant. Yet it also seems to imply and require a cosmological magnetic field to support the Kaluza scalar field. The implied values of $B_c$ seem consistent with intergalactic or primordial magnetic field values above $10^{-10}$ T.

Therefore, $\phi$ is set to $1$ by cosmological parameters. It can be approximately constant over large lengthscales. Variations from the flat-space value are observed around planetary masses and local bulk electromagnetic fields, just as with gravity.  

%%%%%%%%%%%%%%%%%%%%%%%%%%%%%%%%%%%%%%%%%%%%%
%%%%%%%%%%%%%%%%%%%%%%%%%%%%%%%%%%%%%%%%%%
\section{VII. ENERGY CONSIDERATIONS}
%%%%%%%%%%%%%%%%%%%%%%%%%%%%%%%%%%%%%%%%%%%%%
%%%%%%%%%%%%%%%%%%%%%%%%%%%%%%%%%%%%%%%%%%

We have mentioned how Dicke anticipated that any interaction with a scalar field must result in a variation in rest mass. This point was further investigated by Ref.~\cite{frw}. They found that conformal transformations, interactions with a scalar, can be interpreted as ``apparent" gravitational fields because the rest mass includes the potential energy in such a field, just as given in (\ref{U0}). The rest mass of a particle is relative, then. Yet a special value can be singled out in the same way we single out the rest mass associated with the Minkowski metric.

Let us consider another feature of the Kaluza scalar field that was anticipated by Dicke: the action of the scalar field on a body produces acceleration at constant energy. Note the energy is also constant for the Newtonian approximation to gravity. Although the Newtonian gravitational field can accelerate test particles, terms quadratic in speed are ignored, and the energy including rest energy is essentially constant by approximation. 

Energy under action of the Kaluza scalar field is constant for a more fundamental reason. The equation of motion (\ref{4Deom}) shows that if $\p_t\phi =0$, then $dU^t/d\tau = 0$ from the scalar field, and therefore the energy is constant. However, the scalar field is conservative like the gravitational field. Dicke anticipated the energy gained by acceleration under the scalar field would be offset by a loss of rest mass energy from interaction with the scalar field. Here we find the energy of acceleration is offset by a loss of scalar field potential energy. 

Let us consider the 5D energy. Since we are considering the time-independent case, $\p_t {\t g}_{ab} =0$, then the covariant time component ${\t U}_t$ is constant, and given by:
\bea
\label{U0}
{\t U}_t &=& {\rm constant} = g_{tt} {\t U}^t + kA_t\U5 + g_{ti}{\t U}^i\cr
&=& g_{tt}{cd\tau\o ds}{cdt\o d\tau} \\
&\simeq& \left( 1- {GM_{cl}\o rc^2}  + {GM_{cl}\o rc^2}{m_{EG}\o m}\right)
+ \mathscr{O}(\psi^2) + \mathscr{O}(\xi^2)\nonumber
\eea
where we are using the clothed mass $M_{cl}$ of the planet, and where we used (\ref{gradpsi}), (\ref{clth}), (\ref{meg}), and (\ref{emass}), with $cdt/d\tau \simeq 1 $ as usual.

The Kaluza scalar field potential is an additional term in the test particle energy budget, along with rest energy and gravitational potential energy: 
\be
\label{te}
{\rm total\ energy} \rightarrow mc^2 - {GM_{cl}m\o r} + 
{Gm_{EG}M_{cl}\o r}
\ee

An ADM-like analysis \cite{adm} of the Kaluza scalar field potential energy is required to investigate any stabilizing effect on a point particle, such as that found for the electric field. However, we note that (\ref{te}) only applies to macroscopic bodies. Behavior at strong charge states characteristic of elementary particles was considered above, and these are the charge states that would be subjected to an ADM-like analysis.

%%%%%%%%%%%%%%%%%%%%%%%%%%%%%%%%%%%%%%%%%%%%%
%%%%%%%%%%%%%%%%%%%%%%%%%%%%%%%%%%%%%%%%%%
\section{VIII. CONCLUSIONS}
%%%%%%%%%%%%%%%%%%%%%%%%%%%%%%%%%%%%%%%%%%%%%
%%%%%%%%%%%%%%%%%%%%%%%%%%%%%%%%%%%%%%%%%%

\begin{table*}
\caption{\label{tab:table1}Limiting values of scalar, electric, and gravitational potentials \& charges, for bodies of electric charge $Q$ and bare mass $M$; for static fields, $\psi\ll 1$, $\xi\ll 1$, non-relativistic matter, and spherical symmetry. The clothed mass is $4M/3$. Only leading terms in potentials are shown. The term in electric field energy, $M_Ec^2$, is large for ${\t U}_5^2\gg \xi^{-1}$, but is omitted for clarity. $\xi_0 \equiv\xi(r_0) $}
\begin{ruledtabular}
\begin{tabular}{c|cccccc}
%row 1
 Charge state&
 \multicolumn{2}{c}{----------- Scalar -----------}&
 \multicolumn{2}{c}{---------- Electric ----------}&
 \multicolumn{2}{c}{-------- Gravitational -------}\\
%row 2
 (neutral, lab, atomic)&
 charge&
 potential, $\xi$&
 charge&
 potential&
 charge&
 potential, $\psi$\\  \hline \\
%row 3
\(\displaystyle Q=0\)&
 0&
 \(\displaystyle {2\o 3}{GM\o rc^2}\) &
 0&
 0&
 $Mc^2$&
 \(\displaystyle -{8\o 3}{GM\o rc^2}\)\\ \\
%row 4
 \(\displaystyle {{Q^2/M^2}\o G\epsilon_0 } \ll \xi^{-1}\)&
 \(\displaystyle {{c^2 Q^2/M}\o 16\pi G\epsilon_0 } \)&
 \(\displaystyle -{\mu_0\o 12\pi}{Q^2\o Mr} \) &
 Q&
 \(\displaystyle{1\o 4\pi\epsilon_0} {Q\o r}\)&
 $Mc^2$&
 \(\displaystyle {\mu_0\o 12\pi}{Q^2\o Mr} \) \\ \\
%row 5
\(\displaystyle {{Q^2/M^2}\o G\epsilon_0 } \gg \xi^{-1}\)&
 \(\displaystyle {c^2 Q/\sqrt{2\xi}\o \sqrt{16\pi G\epsilon_0 }} \)&
 \(\displaystyle - {\mu_0\over 12\pi}{Q\o r} \sqrt{8\pi G\epsilon_0 \o \xi_0} \) &
 Q&
 \(\displaystyle{1\o 4\pi\epsilon_0} {Q\o r}\)&
 \(\displaystyle {c^2{Q}\sqrt{2\xi}\o \sqrt{16\pi G\epsilon_0 }} \)&
 \(\displaystyle  {\mu_0\over 12\pi}{Q\o r} \sqrt{8\pi G\epsilon_0 \o \xi_0} \) \\ 
%row 6
& & ($\xi\rightarrow 0$ as $r_0\rightarrow 0$) & & & &($\xi\rightarrow 0$ as $r_0\rightarrow 0$)
\end{tabular}
\end{ruledtabular}
\end{table*}

The tensor gravitational potential and the vector electromagnetic potential behave as if they are components of a 5D gravitational potential, but that implies the existence of a third field, a scalar potential in 4D. The Kaluza scalar field is a long range scalar field, presumably associated with a spin 0 massless boson, as the gravitational field is presumably associated with the spin 2 massless graviton.

The absence of a detectable 5th dimension is enforced as a boundary condition on the fields such that derivatives $\p_5 {\t g}_{ab} =0$. Far from being an unnatural or ad hoc simplification, this reveals a non-trivial classical constant of the motion corresponding to electric charge. With this identification, the geodesic equation in 5D provides the 4D gravitational and electromagnetic forces, augmented with a scalar force that couples to scalar charge. In new results here, principles of 5D covariance are found to imply the scalar charge must be proportional to $Q^2/M$. That the scalar charge can be expressed in terms of the gravitational charge (mass) and electric charge is because the 5D length element of a particle is a constant of the motion.

When these considerations are applied to massive charged and neutral bodies, for the simplified cases of static, spherically-symmetric fields and non-relativistic sources, it is found that neutral mass is clothed in the Kaluza scalar field, and so the mass of planets determined from Kepler's laws, and the component $g_{tt}$ of the metric, is a clothed mass that includes contributions from the bare mass and its associated scalar field. The magnitude of this scalar field is the same as calculated in Brans-Dicke theory for $\omega=0$, but the nature of the Kaluza scalar field is much different, in both its couplings and its field equation.

The gravitational, electric, and scalar charges and potentials calculated in this work are tabulated for convenience in Table~\ref{tab:table1}.

The force equations imply that electrically-charged objects immersed in the Kaluza scalar field of planets should experience an electro-gravitic lift that is effectively a buoyancy force. Dicke noted that long range scalar fields have the peculiar property of providing acceleration at constant energy. He anticipated that the energy would come from variation of rest mass. We find acceleration at constant energy here, except the variation in rest mass can be understood as a variation of potential energy in the scalar field, as described by Rohrlich and Witten. The upward momentum provided by the buoyant lifting force is compensated by the ambient scalar field, and by recoil of the field that ultimately couples to recoil of the earth.

The expression (\ref{ksflab}) for the scalar potential when ${\t U}_5 \ll \xi^{-1}$ can be compared with the monopole solution found by Ferrari \cite{fri}. Ferrari found a clever solution by assuming that there were only 2 length scales characterizing an electrically-charged mass: $GM/c^2$ and $Q\sqrt{G/4\pi\epsilon_0 c^4}$. Yet the solution seen in (\ref{ksflab}) indicates there is a third lengthscale, $\mu_0 Q^2 / M$, that is not anticipated by the other two lengthscales from gravity or electric forces. The third lengthscale is an electro-gravitic lengthscale, characteristic of the scalar charge (\ref{schg}). Therefore, the solution by Ferrari, while predicting a significant force from the scalar interaction, does not capture the unique scaling of the scalar interaction. Nor was this lengthscale discerned by Chodos \& Detweiler in their static monopole solution.

Some of the scalar forces predicted for laboratory charges are quite large, and should have been seen already if they exist. Yet it is difficult to tune the theory to drive the scalar forces to zero, without compromising identification with standard physics in other areas. This should therefore be considered the first testable classical falsification or verification of the hypothesis of five-dimensional general relativity.

Principles of 5D covariance imply a saturation effect must exist in the gravitational and electric charges at the high specific charge states characteristic of elementary particles. We find that this saturation effect alters the gravitational and scalar charges such that they go over to dependence on electric charge, possibly masquerading as the electric force in such regimes, before going to zero in the limit of point particles.
%%%%%%%%%%%%%%%%%%%%%%%%%%%%%%%%%%%%%%%%%%%%%
%%%%%%%%%%%%%%%%%%%%%%%%%%%%%%%%%%%%%%%%%%
\section{Acknowledgements}
%%%%%%%%%%%%%%%%%%%%%%%%%%%%%%%%%%%%%%%%%%%%%
%%%%%%%%%%%%%%%%%%%%%%%%%%%%%%%%%%%%%%%%%%

This work was supported by DARPA DSO under award AQD number D19AP00017.

\end{document}